\documentclass{emulateapj}
\usepackage{psfig}

\newcommand{\totder}[2]{\frac{d{#1}}{d{#2}}}
\newcommand{\parder}[2]{\frac{\partial{#1}}{\partial{#2}}}

\newcommand{\gsim}{\mbox{\hspace{.2em}\raisebox{.5ex}{$>$}\hspace{-.8em}\raisebox{-.5ex}{$\sim$}\hspace{.2em}}}
\newcommand{\lsim}{\mbox{\hspace{.2em}\raisebox{.5ex}{$<$}\hspace{-.8em}\raisebox{-.5ex}{$\sim$}\hspace{.2em}}}
\newcommand{\ssst}{\scriptscriptstyle}
\newcommand{\E}[1]{\times 10^{#1}}
\newcommand{\etal}{et al.}
\newcommand{\lt}{\left}       \newcommand{\rt}{\right}
\newcommand{\RA}[3]{{#1}^{{\rm h}}{#2}^{{\rm m}}{#3}^{{\rm s}}}
\newcommand{\Dec}[3]{{#1}^{\circ}{#2}'{#3}''}

\newcommand{\s}{\,{\rm s}}      \newcommand{\ps}{\,{\rm s}^{-1}}
\newcommand{\yr}{\,{\rm yr}}    \newcommand{\Msun}{M_{\odot}}
\newcommand{\cm}{\,{\rm cm}}    \newcommand{\km}{\,{\rm km}}
\newcommand{\parsec}{\,{\rm pc}}
\newcommand{\ergs}{\,{\rm ergs}}        
    \newcommand{\keV}{\,{\rm keV}}
    \newcommand{\G}{\,{\rm G}}
\newcommand{\as}{$^{\prime\prime}\ $} 

\newcommand{\nel}{n_{e}}        \newcommand{\NH}{N_{\ssst\rm H}}

\newcommand{\nH}{n_{\ssst\rm H}}

 \newcommand{\ASCA}{{\sl ASCA}}
\newcommand{\Chandra}{{\sl Chandra}}
	\newcommand{\ru}{R_{12}}
\newcommand{\Bp}{B_{\perp}}

\newcommand{\epm}{\varepsilon_{\rm m}}
\newcommand{\epl}{\varepsilon_{\rm l}}
\newcommand{\epu}{\varepsilon_{\rm u}}

\def\snr{{N157B}}
\def\psr{{PSR J0537$-$6910}}
\slugcomment{To appear in the ApJ 10 October 2006, v650n1 issue}

\shortauthors{Chen et al.}
\shorttitle{\Chandra\ Spectroscopy of SNR N157B}

\begin{document}

\title{{\sl CHANDRA} ACIS Spectroscopy of N157B --- A Young Composite
 Supernova Remnant in a Superbubble}
\author{
 Yang Chen\altaffilmark{1},
 Q.\ Daniel Wang\altaffilmark{2,3},
 E. V. Gotthelf\altaffilmark{4},
 Bing Jiang\altaffilmark{1},
 You-Hua Chu\altaffilmark{5}, and
 Robert Gruendl\altaffilmark{5}
}
\altaffiltext{1}{Department of Astronomy, Nanjing University, Nanjing 210093,
       P.R.China}
\altaffiltext{2}{Department of Astronomy, B619E-LGRT,
       University of Massachusetts, Amherst, MA~01003}
\altaffiltext{3}{Institute for Advanced Study, Einstein Drive, Princeton, NJ 08540}
\altaffiltext{4}{Columbia Astrophysics Laboratory, Columbia University,
       550 West 120 St, New York, NY~10027}
\altaffiltext{5}{Astronomy Department, University of Illinois at
       Urbana-Champaign, 1002 West Green Street, Urbana, IL~61801}

\begin{abstract}
We present a \Chandra\ ACIS observations of N157B, a young supernova
remnant (SNR) located in the 30 Doradus star-formation region of the Large
Magellanic Cloud. This remnant contains the most energetic pulsar
known (J053747.39--691020.2; $\dot E = 4.8 \times 10^{38}$~ergs~s$^{-1}$),
which is
surrounded by a X-ray bright nonthermal nebula that likely represents a
toroidal pulsar wind terminal shock observed edge-on. 
Two of the eight point-like X-ray sources detected in the observation
are shown to have near-IR and optical counterparts (within 0\farcs5 offsets),
which are identified as massive stellar systems in the Cloud.
We confirm the
non-thermal nature of the comet-shaped X-ray emission feature and show
that the spectral steepening of this feature away from the pulsar is
quantitatively consistent with synchrotron cooling of shocked
pulsar wind particles flowing downstream at a bulk velocity close to the
speed of light. Around the cometary nebula we unambiguously detect a
spatially-resolved
thermal component, which accounts for about 1/3 of the total 0.5 -- 10
keV flux from the remnant. This thermal component is distributed among
various clumps of metal-enriched plasma embedded in the low surface
brightness X-ray-emitting diffuse gas.  The relative metal enrichment
pattern suggests that the mass of the supernova progenitor is
$\gsim20\Msun$. A comparison of the X-ray data with {\sl Hubble Space
Telescope} optical images now suggests that the explosion site
is close to a dense cloud, against which a reflection shock is
launched. The interaction between the reflected material and the nebula
has likely produced both its cometary shape and the surrounding
thermal emission enhancement. SNR N157B is apparently expanding into
the hot low-density interior of the surrounding superbubble formed by the
young OB association LH99, as revealed by {\sl Spitzer} mid-infrared
images. This scenario naturally explains the exceptionally large
sizes of both the thermal and nonthermal components as well as the
lack of an outer shell of the SNR.  However, the real situation in the
region is likely to be more complicated. We find that a partially-round-shaped
soft X-ray-emitting clump with distinct spectral properties may result
from a separate oxygen-rich remnant.  These results provide a rare glimpse
into the SNR structure and evolution in a region of recent star-formation.
\end{abstract}


\section{Introduction}

The evolution of a supernova remnant (SNR) depends sensitively on its 
environment. Most massive stars are expected to be born in OB associations 
and end their lives in an environment altered by strong ionizing radiation
and mechanical energy feedback from fast stellar winds and supernovae (SNe).
This feedback produces superbubbles filled with low-density hot gas.  
An SNR inside such a superbubble will be difficult to observe (e.g., 
Tang \& Wang 2005). A classical example is the Crab SNR, which shows 
little detectable emission from its blastwave-heated gas and is believed 
to be expanding in a low-density medium.  Without the synchrotron radiation
from its pulsar wind nebula (PWN), the Crab SNR would not have been 
identified.  Similar SNRs dominated by radiation from PWNe are called 
``Crab-like''.  Physically, Crab-like SNRs are not very different from 
the ``composite'' remnants, which show comparable thermal and nonthermal
components or show a shell-like structure with a centrally-filled morphology.
The increasing sensitivity of X-ray observations has made it possible
to detect more shell-like and thermal features around Crab-like SNRs.
For example, the Crab-like SNR G21.5$-$0.9 has recently been shown to 
be surrounded by ejecta- or blastwave-dominated features
(Slane et al.\ 2000; Matheson \& Safi-Harb 2005).
In this paper, we present evidence that the Crab-like SNR N157B, also 
known as SNR 0538-691, exhibits thermal X-ray emission as well and is 
thus a composite SNR.

SNR N157B is particularly interesting, because of its co-existence
with a young OB association LH99 in the Large Magellanic Cloud (LMC)
(Lucke \& Hodge 1970; Chu et al.\ 1992). The SNR has long been 
classified to be Crab-like and contains the fastest known young pulsar 
\psr~ with a period of 16 ms  (Marshall et al. 1998). The SNR appears
to have an exceptionally large physical size in both radio and X-ray 
($\sim20-30$~pc; Lazendic et al.\ 2000;
Wang \& Gotthelf 1998, WG98 hereafter),
although the outer boundaries have not yet been well determined. Of 
particular interest is how the structure and evolution of SNR N157B is 
affected by the presence of the OB association LH99. Is the SNR 
expanding in a low-density hot medium, as may be expected
in a superbubble created by the OB association? Addressing such 
questions is important in understanding both the fate of the feedback
from massive stars and the physics involved in the evolution of such 
an SNR (e.g., Tang \& Wang 2005).

Previous X-ray studies based on data with low angular resolutions
were concentrated on the overall spectral and morphological 
characterization of \snr.  Using {\it ASCA} and {\it ROSAT} data, 
WG98 found that the X-ray emission arises from a bright cometary-shaped
nebula, headed by a compact source embedded in a bright bar of nonthermal
emission, with a long tail of diffuse emission trailing
on one side aligned perpendicular to the bar.  This morphology is 
reminiscent of those associated with trailing plerions in
SNRs G327.1$-$1.1 (Sun, Wang, \& Chen 1999), W44 (Frail et al.\ 1996),
and IC443 (Olbert et al.\ 2001; Gaensler et al.\ 2006), as well as
the PWN around PSR~B1757$-$24 near SNR~G5.4$-$1.2
(the Duck X-ray Nebula; Kaspi et al.\ 2001).
Using a \Chandra\ HRC observation with high angular and timing
resolutions, Wang et al.\ (2001) were able to pinpoint (to arcsec 
precision) the location of PSR~J0537-6910, which is still
undetected in any other wavelength band
(e.g. Mignani et al. 2005; Crawford et al. 2005).
The nebula was shown to consist of primarily a  $\sim 0.6$ pc 
$\times 1.7$ pc compact bar-like feature around the pulsar and a 
tail $\gsim 5$ pc long. 
This morphology indicates that the bar represents the
reverse shock of a toroidal wind from \psr\
and that the ambient pressure confinement of the nebula is largely one-sided.

\begin{figure*}[thb!] 
\centerline{\hfil\hfil
\psfig{figure=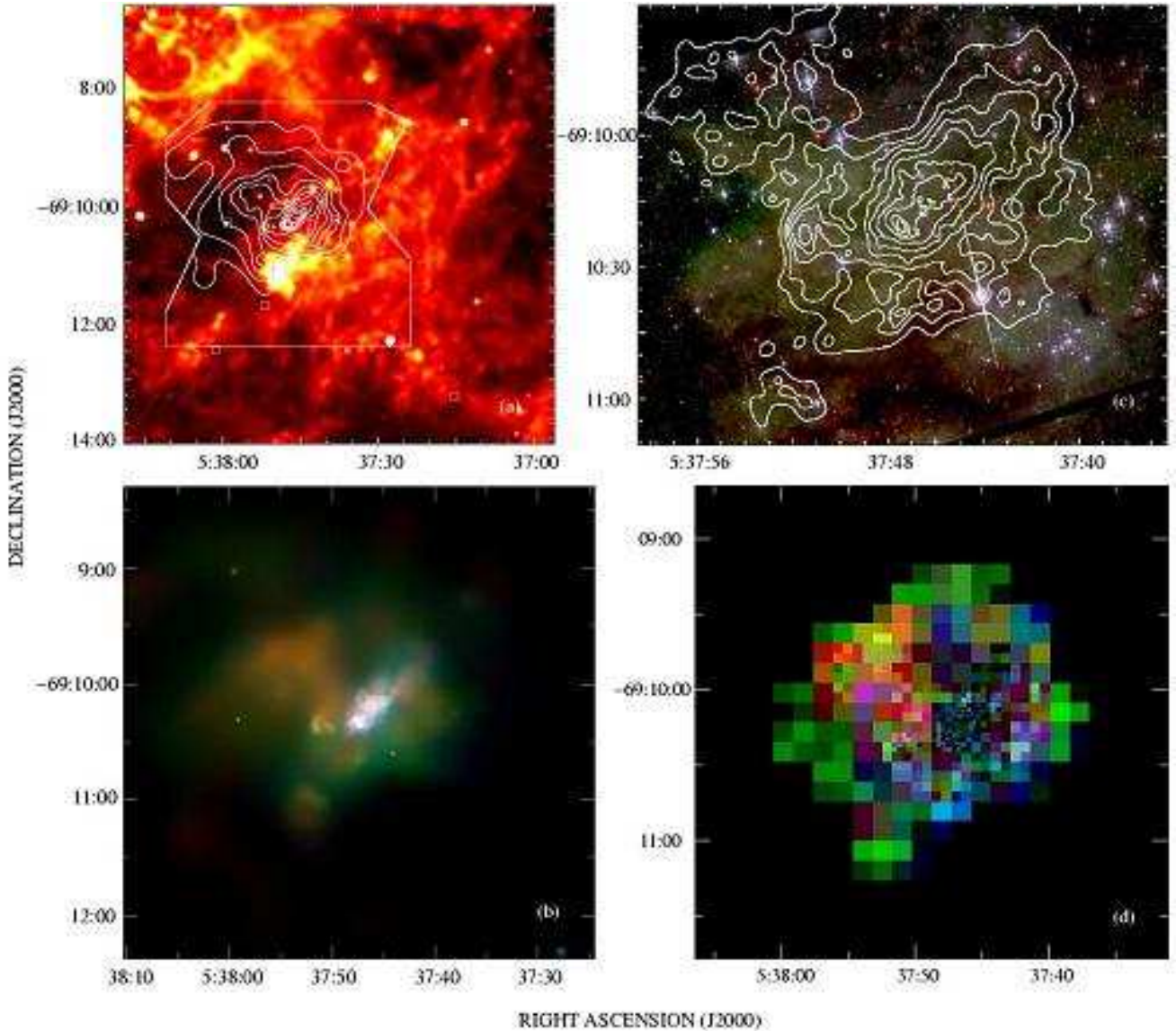,height=6.in,angle=0, clip=}
\hfil\hfil}
\caption{(a) A  {\it Spitzer} global view of the \snr\ region 
in mid-infrared (IR)
(8 $\mu$m; Fazio et al.\ 2004) with overlaid {\sl Chandra} ACIS-S 
intensity contours at 2, 3, 4, 6, 9, 13, 18, 25, 50, 100, 200, 500, and 
1000 $ \times 10^{-2} {\rm~counts~s~^{-1}~arcmin^{-2}}$. The outer
boundary of the contour map shows the region covered in (b). The small 
square boxes mark discrete X-ray sources detected. (b)
Color-coded multi-band ACIS-S intensity images of N157B in the 
0.3-0.7 keV (red), 0.7-1.5 keV (Green), and 1.5-7 keV (blue) bands.
(c) ACIS-S intensity contours at 12, 20, 30, 50, 100, 250, 800, 2500,
10000 $\times 10^{-2}$ counts~s$^{-1}$~arcmin$^{-2}$
overlaid on an {\sl HST} ACS close-up of the \snr\ central region 
in optical: F814W (color-coded in red), F555W (green), 
and F435W (blue). (d) Equivalent width (EW) images of the 
emission lines: O (in red), Ne (green),
and Mg (blue) in square-root brightness scales;
the line energies are listed in Table~\ref{T:EW}
and the image for each band is adaptively rebinned to include
at least 10 counts per pixel.  
The X-ray images in (a) and (b) are smoothed with the {\small CIAO} routine
{\sl csmooth}, whereas the image in (c)
is smoothed with a Gaussian kernel adjusted adaptively to achieve
a count-to-noise ratio of 6.
}
\label{f:im_color}
\end{figure*}

The one-sided confinement of the \snr\ nebula is thought to be a result of 
the pulsar motion relative to the ambient medium. 
This hypothesis seems to be consistent with the displacement of the pulsar
from an extended radio continuum emission peak, mostly representing 
the accumulation of pulsar wind particles, assumed to be mainly
electrons and positrons (WG98). 
Based on \hbox{2-D} hydrodynamic simulations,
van der Swaluw, Downes, \& Keegan (2004) and van der Swaluw (2004) 
have concluded that a typical PWN evolves 
through three major stages: 1) expanding supersonically in freely 
expanding SN ejecta; 2) interacting with the reverse shock of the SNR, 
oscillating and relaxing to a subsonic expansion in the reverse-shock 
heated ejecta; and 3) heading into the general ISM, supersonically. In both
the first and the third stages, one may expect a bow shock because of 
supersonic motion. They suggest that the PWN of the \snr\ nebula
is currently in the second stage;
the tail of the nebula is a pulsar wind bubble that was elongated owing
to the pulsar motion and was later crushed by the reverse shock
of the SNR. The reverse shock heats a shell of ejecta swept-up by the
expansion of the PWN and bounces back into the already heated remnant
materials, forming a parabolic layer (see Fig.~2 in van der Swaluw 2004).
In this scenario, the relative motion between the pulsar and its ambient
medium does not need to be supersonic. 
Of course, 
the reality could be much more complicated than what is assumed in 
the simulations. The region shows clearly large density and temperature 
inhomogeneities that could have significantly affected the dynamics of 
both the SNR and the PWN (Fig.~\ref{f:im_color}a,b). In short, 
the exact nature of the confinement as well as the physical properties
of the nebula are still very uncertain.  

Existing studies have also shown evidence for a 
weak thermal emission component of \snr. Much of the evidence comes from
the spectral decomposition of the overall emission from the SNR. The thermal
emission arises over a region more extended than the PWN, but
only very limited spatial analysis has been presented previously
(WG98; Dennerl et al.\ 2001;
Townsley et al. 2006). It is thus of great interest to examine the physical, 
chemical, and morphological properties of the thermal emission and to 
determine how they are related to the PWN and to the OB association and 
superbubble environment.

\begin{figure}[tbh!] 
\centerline{ {\hfil\hfil
\psfig{figure=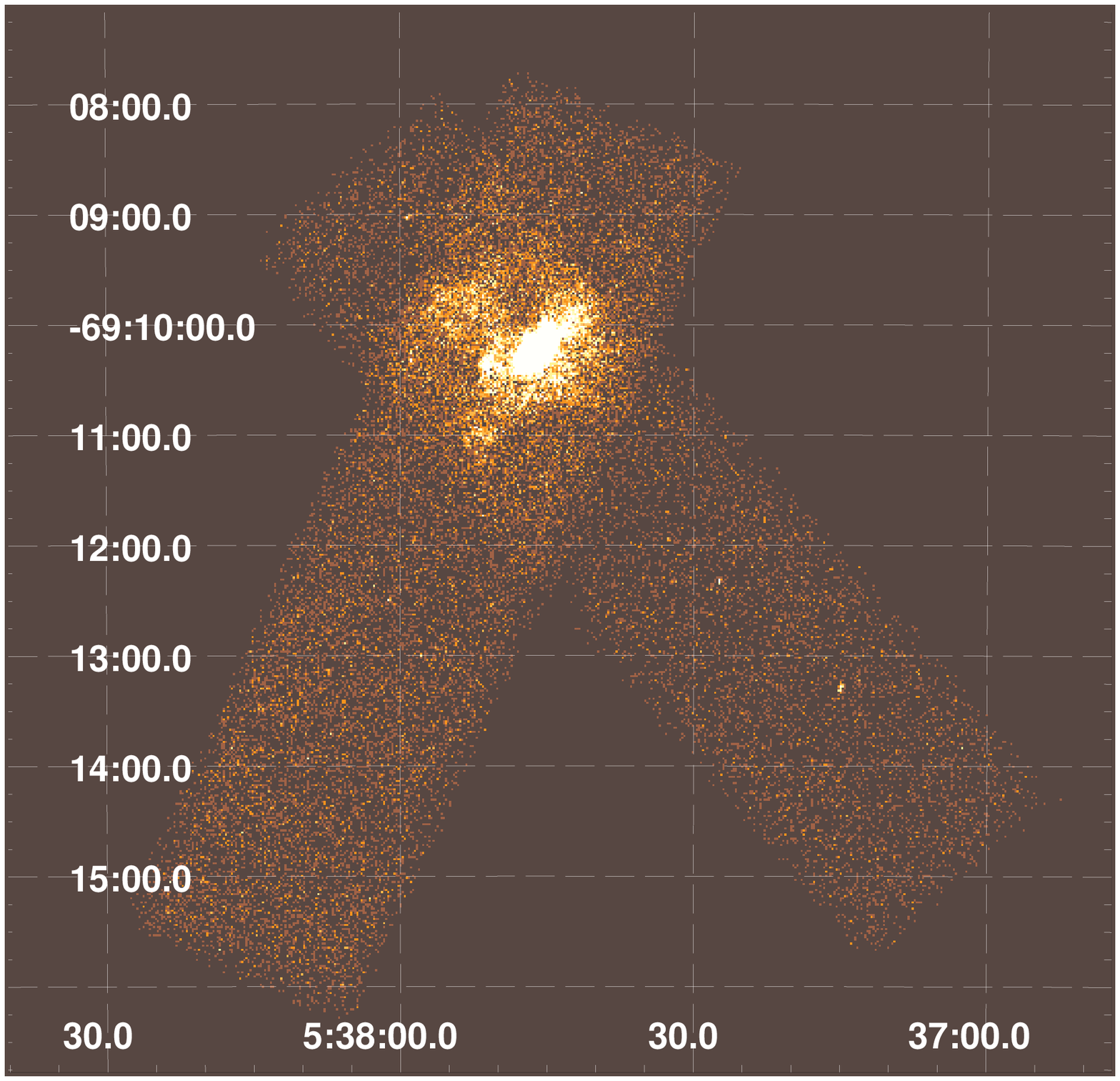,height=3.0in,angle=0, clip=}
\hfil\hfil}}
\caption{The raw data image of the {\sl Chandra} ACIS-S 
observation of SNR N157B.
The coordinate axes are for R.A.\ (J2000) and Dec.\ (J2000), respectively.}
\label{f:acis_all}
\end{figure}

To obtain a fresh look of the above issues, we have acquired an on-axis 
\Chandra\ ACIS-S observation of \snr\ (Fig.~\ref{f:acis_all}), 
allowing for a detailed 
spatially-resolved X-ray spectroscopy of the SNR. We have also analyzed the
X-ray data together with complementary archival images from recent 
{\sl Hubble Space Telescope} ({\sl HST}) Advanced 
Camera for Surveys (ACS)  and 
{\sl Spitzer Space Telescope} Infrared Array Camera (IRAC) observations.
In \S~2, we briefly describe
the observation and data calibration, and present our analysis and
results. The effects of the superbubble and density
inhomogeneity, the origin of the thermal emission, and the nature of
various nonthermal features of the PWN are discussed in \S~3. We
summarize our results and conclusions in \S~4. Throughout this paper,
we adopt the distance of N157B as 50 kpc (thus $1^{\prime\prime}$
corresponds to 0.25 pc). Statistical errors are all presented in the
90\% confidence level.

\section {Observations and Data Analysis}

The X-ray study reported here is based on our {\sl Chandra}
observation of N157B (ObsID 2783). The data were collected in two
exposures, obtained on 2002 August 23--24 and on 2002 October 28--30,
for a combined duration of 53 ks. The data from the two exposures were
merged through the pipeline processing and no additional alignment
is attempted.  Because of the separation in time,
the relative roll angle of these exposures is $65\deg$ 
(Fig.~\ref{f:acis_all}). Data were
collected with ACIS-S S3 in the focal plane using the 1/4-chip
sub-array mode (with a frame readout time of 0.8 s) to minimize photon
pile-up associated with the bright pulsar. The
first part of the observation (30 ks exposure) has been used by
Mignani et al. (2005) in a primarily optical study of \psr, including
a search for its optical counterpart and an X-ray spectral analysis.

We reprocess the event files (from Level 1 to Level 2) for the
observation using the updated {\small CIAO} data processing software. 
After removing flares with count rates greater than
1.2 times the mean light curve value, a net exposure of 48 ks remains 
and is used for subsequent analysis.  For the spectral analysis, 
individual spectra are adaptively binned to achieve a 
background-subtracted signal-to-noise ratio of 5, unless 
otherwise noted. The {\small XSPEC} spectral fitting software is used
throughout.  For the foreground absorption, we use the cross-sections
of Morrison \& McCammon (1983) and assume solar abundances.

\begin{figure*}[tbh!] 
\centerline{ {\hfil\hfil
\psfig{figure=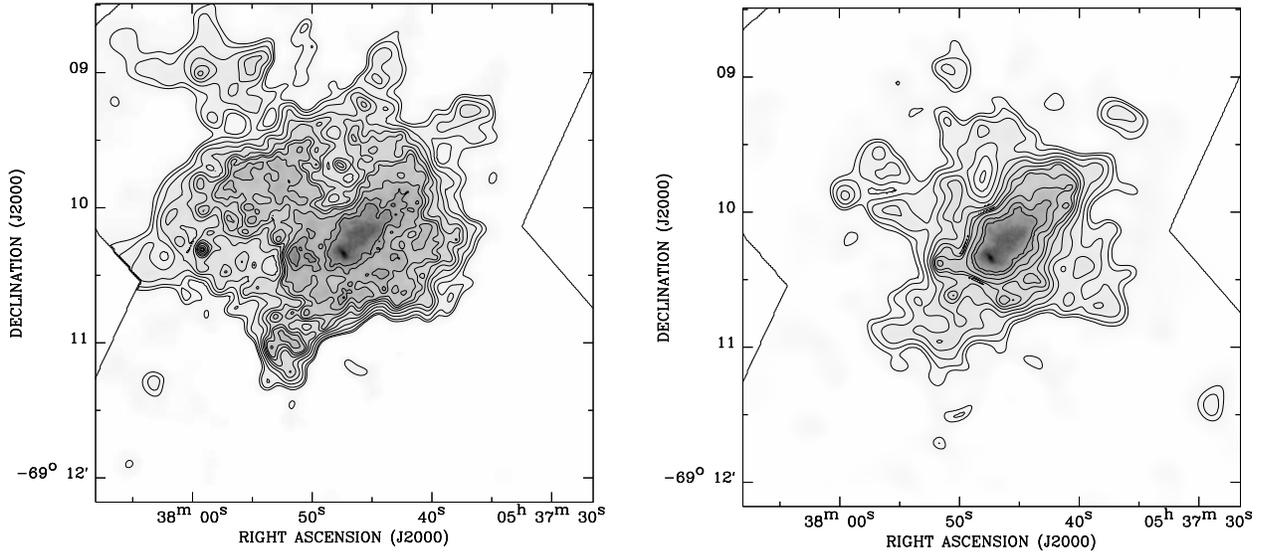,height=3.5in,angle=90, clip=}
\hfil\hfil}}
\caption{{\sl Chandra} ACIS-S images of SNR N157B in the 0.3 -- 1.5 keV
band (left panel) and the 1.5 -- 7 keV band. The images are smoothed with 
a Gaussian with its size adjusted adaptively to achieve a count-to-noise 
ratio of 6. The contours are at 3, 4, 6, 9, 13, 18, 24, 31, 50, 100, and 
200$\sigma$ above a background of 1.5 $\times 10^{-2} 
{\rm~counts~s~^{-1}~arcmin^{-2}}$. 
The outer boundaries of the sub-array boundaries  are outlined.}
\label{f:2band}
\end{figure*}

Fig.~\ref{f:2band} presents the overall \Chandra\ X-ray intensity
image in the soft ($0.3-1.5\keV$) and hard ($1.5-7.0\keV$) energy bands, 
which clearly demonstrate the energy dependence of the SNR morphology.
The images do not show a typical rim-brightened outer SNR shell; 
therefore, the overall size of the diffuse X-ray emission is 
uncertain. Here we will adopt a diameter of the remnant as $\sim100''$.
The diffuse emission also exhibits various lumpy sub-structures.  
Intensity maps and corresponding exposure maps were also generated 
in four broad energy bands, 0.3 -- 0.7 keV, 0.7 -- 1.5 keV, 
1.5 -- 3.0 keV, and 3.0 -- 7.0 keV (Wang 2004). These maps are used to 
create the flat-fielded count rate images, which are typically 
smoothed with an adaptive filter (using the program {\em csmooth} 
implemented in the {\small CIAO} software package) to achieve a 
broad-band (0.3 -7 keV) signal-to-noise ratio of $\sim 3$.  

To compare the distribution of the X-ray emission with its interstellar
environment of N157B, we have used {\sl HST} ACS/WFC
and {\sl Spitzer} IRAC images.  
The pipeline-processed  ACS images from {\sl HST} program 9471 (PI: Mignani), 
taken with the F435W, F555W, and F814W filters,
were retrieved from the MAST archive 
and are presented without further processing.  
{\sl Spitzer} observed the 30\,Dor region with the IRAC (Fazio et al.\ 2004)
as part of programs 63, 1032, and 20203 (PIs: Houck, Brandl, and 
Meixner, respectively).
Each program used the IRAC mapping mode to obtain maps of the region with 
multiple dithered 12~s exposures at each location in the map.  
The Basic Calibrated Data (BCDs) from these programs were combined using 
standard routines in the MOPEX software package to obtain images with 
$\sim$180~s integration time at all locations where the programs overlap.
More information on the instruments, pipeline processing, and the  
MOPEX package can be found at the {\sl Spitzer} Science Center's 
Observer Support website\footnote{http://ssc.spitzer.caltech.edu/ost.}.

Fig.~\ref{f:im_color} presents various perspectives of SNR~N157B and
its environment. The X-ray photon energy dependence is illustrated in
tricolor images, for broad-band intensities and for equivalent widths
(EWs) of key emission lines listed in Table~1.
The comet-shaped nebula, as presented in
WG98 and Wang et al.\ (2001), stands out as the brightest and hardest
X-ray feature in the field. Surrounding the nebula is a large-scale
diffuse emission with relatively low-surface brightness and soft
spectral characteristics.  These smoothed X-ray images are
also compared with both optical and mid-IR maps to illustrate their
complex relationship.

\begin{center}
\begin{deluxetable}{lccc}
\tabletypesize{\footnotesize}
\tablecaption{Energy Bounds of the EW images}
\smallskip
\tablewidth{0pt}
\tablehead{
\colhead{Elements} & \colhead{Line}
 & \colhead{Low$^{\rm a}$} & \colhead{High$^{\rm a}$}\\
\colhead{} & \colhead{(eV)} & \colhead{(eV)}& \colhead{(eV)}
}
\startdata
O & 590--740 & 300--550 & 1160--1260\\
Ne & 870--1140 & 750--870 & 1160--1260\\
Mg & 1290--1430 & 1160--1260 & 1620--1720
\enddata
 \tablenotetext{a}{\phantom{0} The low- and high-energy ranges around
the selected line energies used to estimate the underlying continuum.}
 \label{T:EW}
\end{deluxetable}
\end{center}

Point sources in the field were detected following the procedure
detailed in Wang (2004) using a combination of algorithms: wavelet,
sliding box, and maximum likelihood centroid fitting in the above
energy bands.  The source count rate is estimated based on the net
counts within the 90\% energy-encircled radius (EER) of the
telescope/ACIS point-spread function (PSF; Jerius et al.\ 2000).
Table~\ref{acis_source_list} lists our detected point-like sources
whose positions are marked in Fig.~\ref{f:im_color}a.
Because of the exposure correction for the exposure is very small in
the subarry (smaller than the systematic errors involved in
calculating the count rates), the actual count rates without
correction are included in the table.

We search for possible counterparts in the SIMBAD, NED, and Two Micron
All Sky Survey (2MASS) databases (Skrutskie et al.\ 2006) as well as
in various optical images.
We find two optical stellar counterparts, 1-28 (type O5 f*, V=14.26)
and 1-98 [O5 V((f)), V=13.92; Schild \& Testor 1992],
for the X-ray sources \# 3 and 7, respectively (Table~\ref{acis_source_list}).
The most accurate (0\farcs1 RMS) positions of these two stars are
obtained from 2MASS: R.A.\ (J2000), Dec.\ (J2000) =
$5^{\rm h}37^{\rm m}37\fs97,
-69^\circ10^\prime 14\farcs7$ for 1-28 and $5^{\rm h}37^{\rm m}59\fs43,
-69^\circ9^\prime 1\farcs1$ for 1-98; the offsets from the respective
X-ray sources are 0\farcs1 and 0\farcs5, consistent with their position
uncertainties (Table~\ref{acis_source_list}). Therefore, the absolute
astrometry accuracy of the X-ray observation is likely better than 0\farcs5.
Except for the pulsar (\# 4), the other detected X-ray sources are all
quite faint.  Assuming the X-ray spectrum as a thermal plasma with a
temperature of 0.6 keV (for soft case) and 2 keV (for hard case),
respectively, typical for normal early-type stars, and a foreground
absorption of $\NH \sim 6 \times 10^{21} {\rm~cm^{-2}}$ in the region,
we estimate the 0.3--8 keV luminosities of the two X-ray sources
as $\sim (1.2$--$3.6) \E{33}\ergs\ps$ based on the count rates in
Table~\ref{acis_source_list}. These luminosities are higher than
$10^{33}\ergs\ps$, indicating that the X-ray sources most
probably represent massive colliding stellar wind binaries, while
they may still possibly be individual early O-type or WR stars
(e.g., Oskinova 2005). The source \# 6, very soft and without an optical
counterpart, may be a foreground cataclysmic variable. Other sources
are at larger off-pulsar distances and are probably also not related to
N157B.


\begin{center}
\begin{deluxetable*}{lccrrrrrr}
  \tabletypesize{\footnotesize}
  \tablecaption{{\sl Chandra} Source List \label{acis_source_list}}
\tablewidth{0pt}
  \tablehead{
  \colhead{Source} &
  \colhead{CXOU Name} &
  \colhead{$\delta_x$ ($''$)} &
  \colhead{CR $({\rm~cts~ks}^{-1})$} &
  \colhead{HR} &
  \colhead{HR1} &
  \colhead{Flag} \\
  \colhead{(1)} &
  \colhead{(2)} &
  \colhead{(3)} &
  \colhead{(4)} &
  \colhead{(5)} &
  \colhead{(6)} &
  \colhead{(7)}
  }
  \startdata
   1 &  J053715.18--691316.6 &  0.7 &$     2.49  \pm   0.40$& $ 0.67\pm0.13$ &--& B \\
   2 &  J053727.56--691218.4 &  0.6 &$     1.22  \pm   0.27$& --& --& B \\
   3 &  J053737.97--691014.8 &  0.4 &$     0.32  \pm   0.10$& --& --& B \\
   4 &  J053747.39--691020.2 &  0.2 &$   205.83  \pm   2.30$& $ 0.36\pm0.01$ & $ 0.92\pm0.01$ &H \\
   5 &  J053751.79--691143.3 &  0.7 &$     0.30  \pm   0.11$& --& --& H \\
   6 &  J053759.15--691018.4 &  0.3 &$     0.83  \pm   0.15$& $-0.89\pm0.13$ & $ 0.97\pm0.06$ &B \\
   7 &  J053759.48--690901.6 &  0.4 &$     0.62  \pm   0.20$& --& --& B \\
   8 &  J053801.42--691229.1 &  0.6 &$     0.74  \pm   0.18$& --& --& B 
\enddata
\tablecomments{The definition of the bands:
0.3--0.7 (S1), 0.7--1.5 (S2), 1.5--3 (H1), and 3--7~keV (H2). 
In addition, S=S1+S2, H=H1+H2, and B=S+H.
 Column (1): Generic source number. (2):
{\sl Chandra} X-ray Observatory (unregistered) source name, following the
{\sl Chandra} naming convention and the IAU Recommendation for Nomenclature
(e.g., http://cdsweb.u-strasbg.fr/iau-spec.html). (3): Position
uncertainty, including an 1$\sigma$ statistical error calculated from
the maximum likelihood centroiding and an approximate off-axis angle ($r$)
dependent systematic error $0\farcs2+1\farcs4(r/8^\prime)^2$
(an approximation to Fig.~4 of Feigelson et al. 2002), which are added in
quadrature. 
(4): On-axis source broad-band count rate --- the sum of the
exposure-corrected count rates in the four
bands. (5-6): The hardness ratios defined as
${\rm HR}=({\rm H-S2})/({\rm H+S2})$, and ${\rm HR1}=({\rm S2-S1})/{\rm S}$, 
listed only for values with uncertainties less than 0.2.
(7): The labels ``B'', ``S'', or ``H'' mark the bands in
which a source is detected with the most accurate position that is adopted in
Column (2).
}
  \end{deluxetable*}
\end{center}

\begin{figure}[tbh!] 
\centerline{ {\hfil\hfil
\psfig{figure=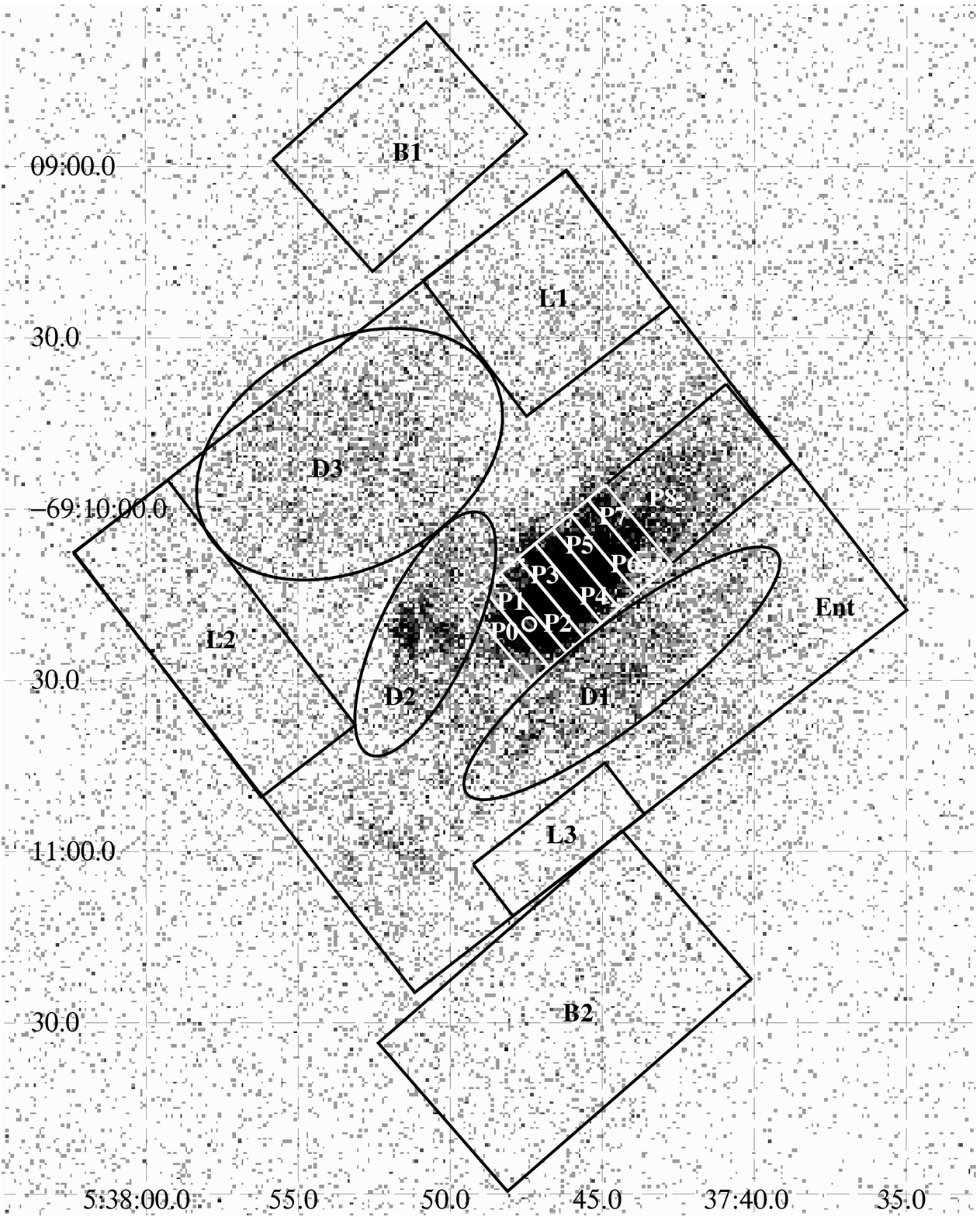,height=4.2in,angle=0, clip=}
\hfil\hfil}}
\caption{Raw ACIS-S image of N157B. The labeled regions are used for
spectrum extraction. The small circle in Region P1 encircles the emission
of \psr, and other point sources are removed.
The coordinate axes are for R.A.\ (J2000) and Dec.\ (J2000), respectively.}
\label{f:reg}
\end{figure}

\subsection{Spectral Properties}\label{sec:glo}

\begin{figure}[tbh!] 
\centerline{ {\hfil\hfil
\psfig{figure=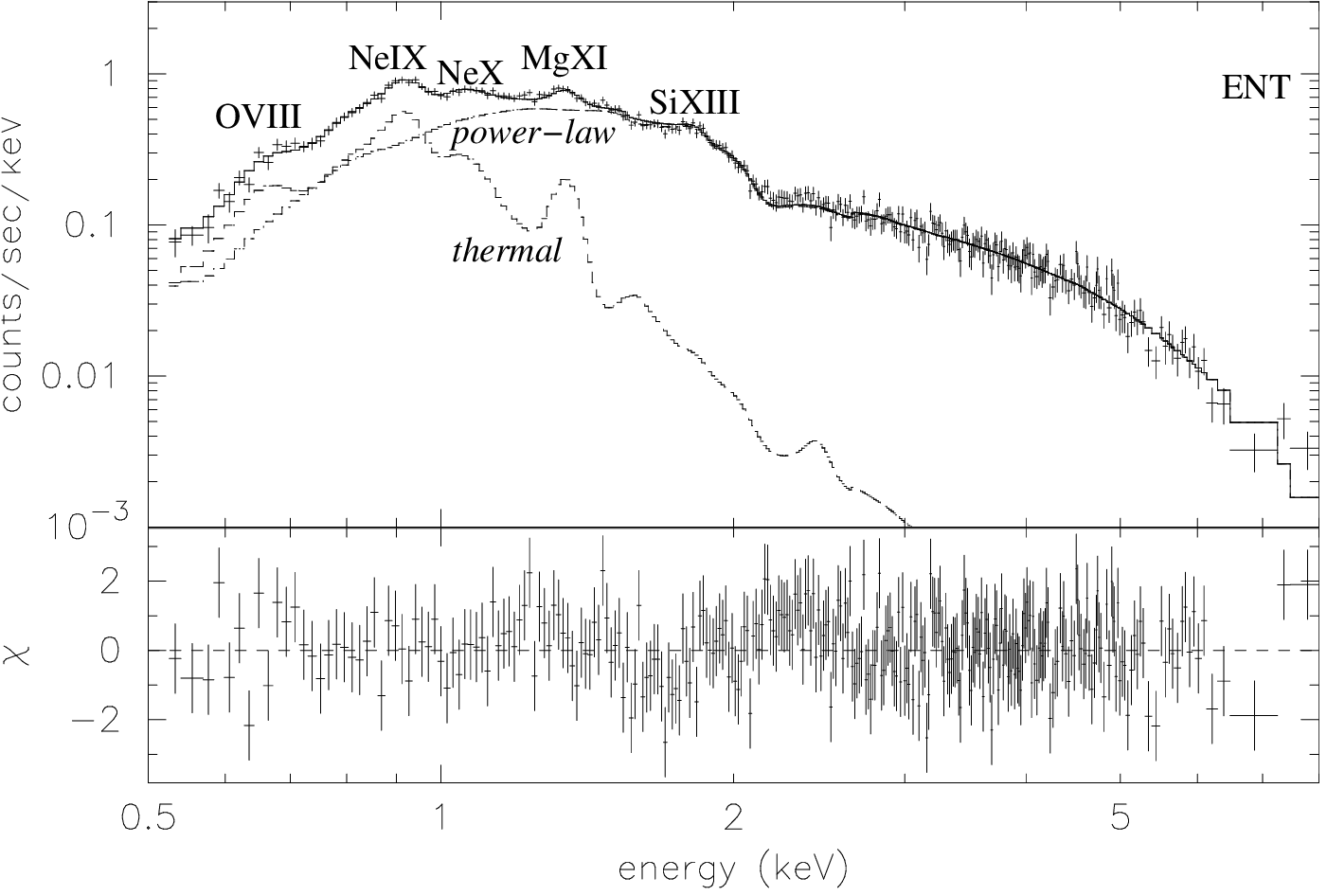,height=2.3in,angle=270, clip=}
\hfil\hfil}}
\caption{The overall spectrum of SNR N157B fitted with the model of
a power-law plus an NEI thermal plasma. The lower panel (likewise also in
other spectral plots) shows the
residuals of the data minus the model in units of $\sigma$.
The two spectral components are
also shown separately.}
\label{f:spec_ent}
\end{figure}

We extracted spectra for analysis from the various features marked in
the raw count image shown in Fig.~\ref{f:reg}. This figure shows some
details that have been smoothed out in the previous images.  The
overall spectrum of the SNR is taken from a large rectangular region
(labeled ``ENT'' in Fig.~\ref{f:reg}), including the pulsar. The
background spectrum is extracted from the southern and northern
rectangular regions (B1 and B2) marked in Fig.~\ref{f:reg}.
Because of the uncertainty of the actual extent of SNR,
the background regions (B1 and B2) may still
be contaminated by the diffuse emission from the remnant. The
background-subtracted spectrum is shown in Fig.~\ref{f:spec_ent}.

Though nearly featureless above $\sim 2\keV$, the spectrum
shows evidence for the K$\alpha$ emission lines of
O~VIII ($\sim0.65\keV$), Ne~IX ($\sim0.92\keV$)
and Ne~X ($\sim1.05\keV$), Mg~XI ($\sim1.34\keV$), 
and possibly Si~XIII ($\sim1.87\keV$).
We fit the spectrum with a power law plus a non-equilibrium
ionization (NEI) thermal plasma that characterizes the parameters and
ionization timescale ($n_e t_i$) of the hot gas
(Borkowski, Lyerly, \& Reynolds 2001). Our fit allows 
the abundances of the above significant line-emitting elements 
to vary, while other metal elements are set to be 0.3 solar,
typical for the LMC (Russel \& Dopita 1992).

The fit results are summarized in Table~\ref{T:ent}.
The luminosity of the nonthermal  0.5--10 keV emission,
including the contributions from \psr\ and the PWN, is
$L_{\rm nt}\sim4.4\E{36}\ergs~\ps$, twice that of the thermal emission
$L_{\rm th}\sim2.2\E{36}\ergs~\ps$.
Clearly, the overall spectrum of \snr\ is 
dominated by the power-law emission, particularly at high energies
($\gtrsim 2$ keV).
As a comparison, the overall luminosity of N157B is independently
estimated as $4.2\E{36}\ergs~\ps$ (0.5 -- 8 keV) from the \Chandra\
ACIS-I observation applying a collisional ionization model to
the thermal component (Townsley et al.\ 2006).
The thermal component is found to be neon-rich and is also slightly
over-abundant in oxygen and magnesium, in respect to 
LMC values, indicating that much of the thermal
emission arises from the SN ejecta.

\begin{center}
\begin{deluxetable}{ll} 
\tabletypesize{\footnotesize}
\tablecaption{Spectral Fit for the Entire SNR}
\smallskip
\tablewidth{0pt}
\tablehead{
\colhead{parameter} & \colhead{value}
}
\startdata
$\chi^2$/d.o.f. & 299.2/301 \\
$\NH$ ($10^{21}\cm^{-2}$) & $6.1\pm0.6$ \\
\span {\em power-law component} \\
$\Gamma$ & $2.29^{+0.05}_{-0.06}$ \\
flux ($10^{-11}\ergs\cm^{-2}\ps$) & $1.4$ \\
\span {\em NEI thermal component}\\
$kT$ (keV) & $0.72^{+0.13}_{-0.10}$ \\
{[O/H]} & $0.43^{+0.16}_{-0.13}$\\
{[Ne/H]} & $1.21^{+0.44}_{-0.13}$\\
{[Mg/H]} & $0.76^{+0.37}_{-0.19}$\\
{[Si/H]} & $0^{+0.40}_{-0}$\\
$n_e t_i$ ($10^{10}\cm^{-3}\s$) & $2.39^{+3.05}_{-1.16}$ \\
$fn_e\nH V^{\rm a}$ ($10^{58}\cm^{-3}$) & $5.22^{+1.49}_{-1.95}$ \\
flux ($10^{-11}\ergs\cm^{-2}\ps$) & $0.7$
\tablecomments{The unabsorbed fluxes are in the 0.5--$10\keV$ band.
The net count rate is $1.090\pm0.005$ counts$\ps$.}
\enddata
\tablenotetext{a}{\phantom{0} Quantity $f\nel\nH V$ is the volume emission
 measure of the hot gas, where $f$ denotes the filling factor.}
\label{T:ent}
\end{deluxetable}
\end{center}

To see where the metal-rich materials are located, we present
the EW images of the O, Ne, and Mg lines
in Fig.~\ref{f:im_color}d. These images are constructed 
in a method similar to those used by Hwang et al.\ (2000) and Park et al.\ (2002).
A major difference of our method from theirs 
is that we rebin the data using an adaptive mesh with each
bin including at least 10 counts
(see Warren et al.\ 2003 for similar binning).
The figure shows that
the line emission is distributed over a broad region, except for the cometary 
X-ray nebula, which is dominated by nonthermal emission and appears 
featureless. The Ne emission (in green) has the largest 
extent, while the Mg line (in blue) is strong 
mostly near the southwestern edge of the nebula (Region D1), while
the O line (in red) is concentrated in a bright northeast patch (D3).

\begin{figure}[tbh!] 
\centerline{ {\hfil\hfil
\psfig{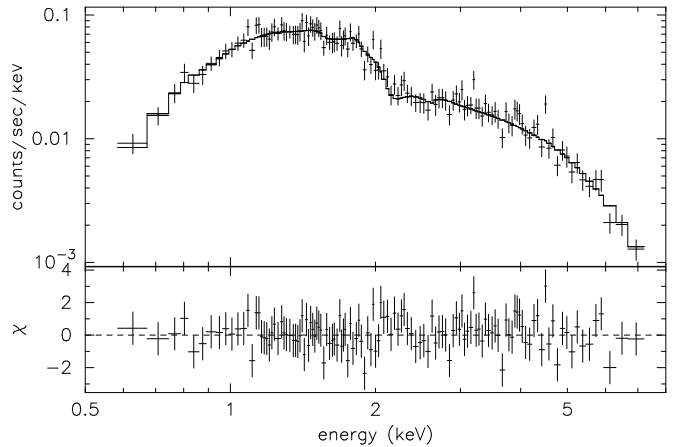}
\hfil\hfil}}
\caption{ACIS-S spectrum of \psr\ and the best-fit power-law.
}
\label{f:spec_p}
\end{figure}

\subsection{\psr}\label{sec:psr}

The location of \psr\ corresponds to the emission peak of the \snr\ 
nebula. The position listed in Table~\ref{acis_source_list} is consistent
with the time-resolved peak of the pulsed emission from the  pulsar 
($\RA{5}{37}{47}.36$, $\Dec{69}{10}{20}.4$;
Wang et al. 2001; see also Mignani et al. 2005). We extract
a spectrum of \psr\ or the core of the nebula bar from a circle 
of $1''$ radius (90\% EER) centered on the emission peak 
and a background from the bar minus the circle (Region P1 in 
Fig.~\ref{f:reg}).
We fit the spectrum
with a power-law model (Fig.~\ref{f:spec_p}; 
no evidence is found for any significant pile-up, based on the model 
given by Davis 2001). The fit is satisfactory 
($\chi^2/{\rm d.o.f.} =  110.0/125$) and gives the best-fit power law index 
as $\Gamma = 1.73^{+0.11}_{-0.06}$ and the foreground
absorption column density as $\NH = 5.6^{+0.5}_{-0.3} \times 
10^{21}\cm^{-2}$. The absorption-corrected 
flux is $\sim1.9\E{-12}\ergs\cm^{-2}\ps$ in the 0.5-10 keV band 
(or a luminosity of $\sim5.7\E{35}\ergs~\ps$). The results are consistent 
with the preliminary results given by Mignani et al.\ (2005).
The power law index is also consistent with that measured for the
pulsed X-ray emission from the pulsar ($\sim1.6^{+0.4}_{-0.3}$),
based on the \ASCA\ and {\sl RXTE} spectra (Marshall et al.\ 1998).
Assuming the same power law ($\Gamma = 1.73$), we find that the
pulsed fraction is 47\% in the 0.5--10 keV band, compared to 
40\% in the 0.2--4 keV band (Wang et al.\ 2001).

In the {\sl Spitzer} IRAC images there are
no point sources at the location of the pulsar.   
The upper limits for the detection of a point source with 
3$\sigma$ confidence are flux densities of 22, 28, 53, 
and 143 $\mu$Jy in the IRAC 3.6, 4.5, 5.8, and 8.0 $\mu$m bands, 
respectively. Similarly, no radio or optical counterpart has been
detected for the pulsar (Crawford et al.\ 2005; Mignani et al.\ 2005).

\begin{figure*}[tbh!] 
\centerline{ {\hfil\hfil
\psfig{figure=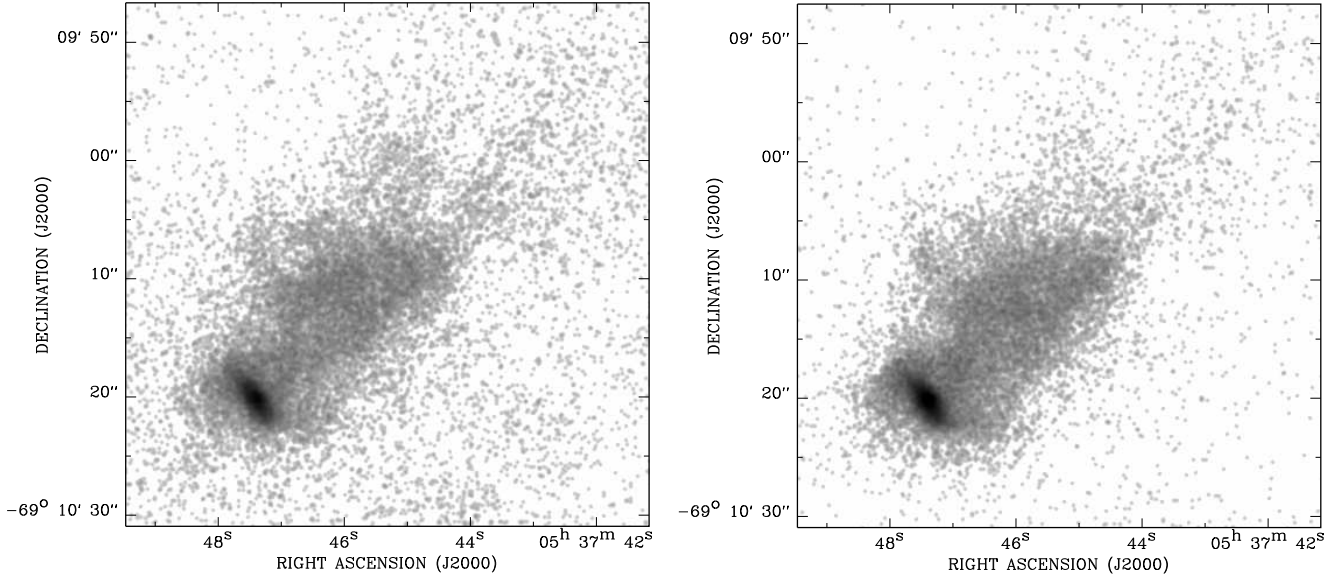,height=3.2in,angle=0, clip=}\hspace{2mm}
\hfil\hfil}}
\caption{Close-ups of the PWN in the 0.3--1.5 keV band (a)
and the 1.5--7 keV band (b). 
These raw count distributions, smoothed with a Gaussian with FWHM=0\farcs3,
are plotted logarithmically.}
\label{f:closeup}
\end{figure*}

\subsection{Diffuse Emission: Nonthermal Component}

Our analysis of the global diffuse X-ray distribution shows that
the cometary nebula has a hard and nearly featureless spectral
characteristics, confirming the conjecture made in the previous
studies (e.g., WG98).  Fig.~\ref{f:closeup} further shows some
interesting details that have not clearly been identified before.  
These include evidence of a halo around the bar that is most
apparent in front of the bar and extends about 5\as from the pulsar.
In the 0.3--1.5 keV band, the new data also show considerable
substructures in the tail.
However, the counting statistics of the data does
not allow for a quantitative characterization of these substructures.

\begin{figure}[tbh!] 
\centerline{ {\hfil\hfil
\psfig{figure=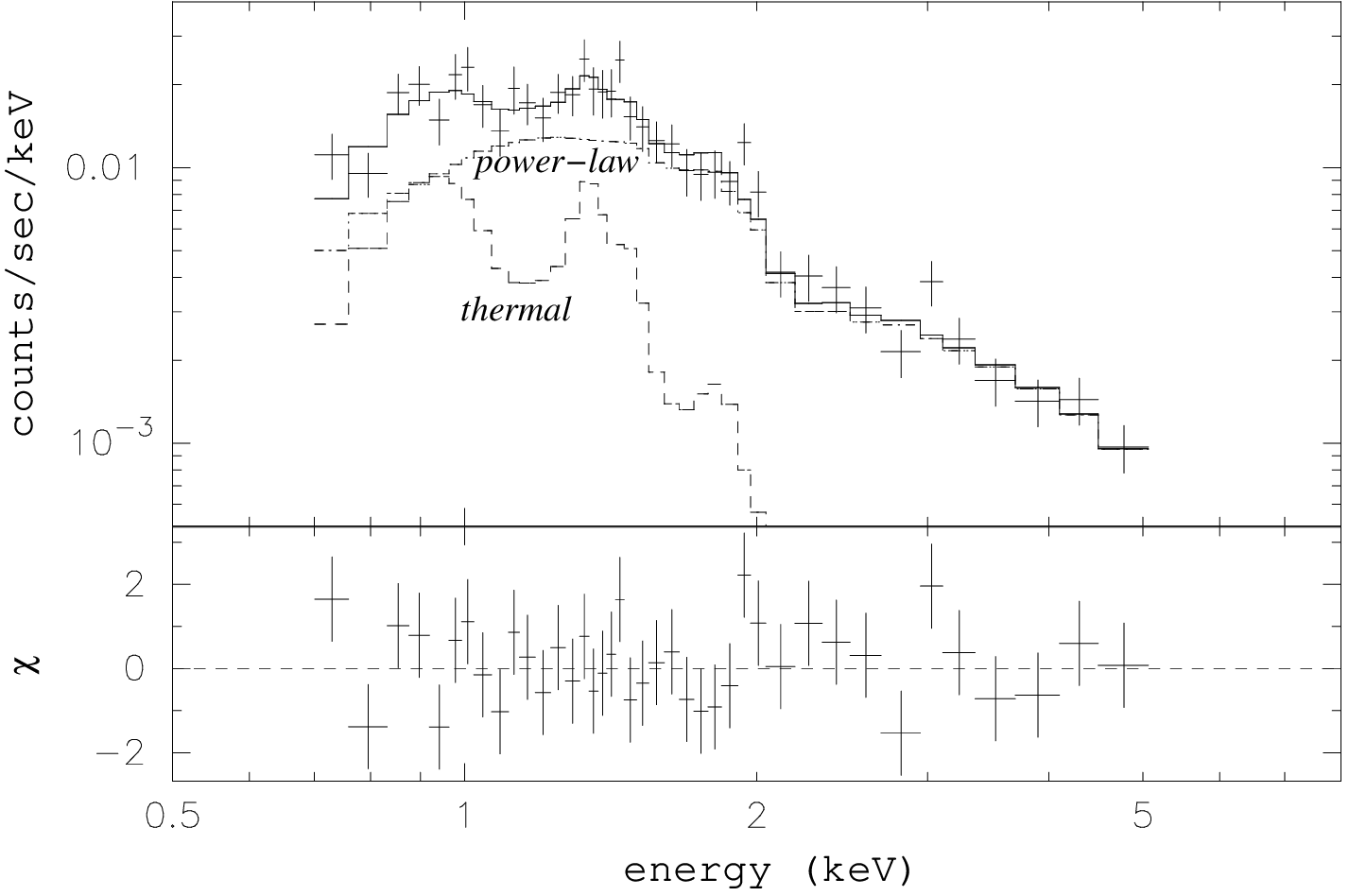,height=2.3in,angle=270, clip=}
\hfil\hfil}}
\caption{ACIS-S spectrum in the regions P0, fitted with 
a power-law plus an NEI thermal plasma.}
\label{f:spec_p0}
\end{figure}

\begin{figure}[tbh!] 
\centerline{ {\hfil\hfil
\psfig{figure=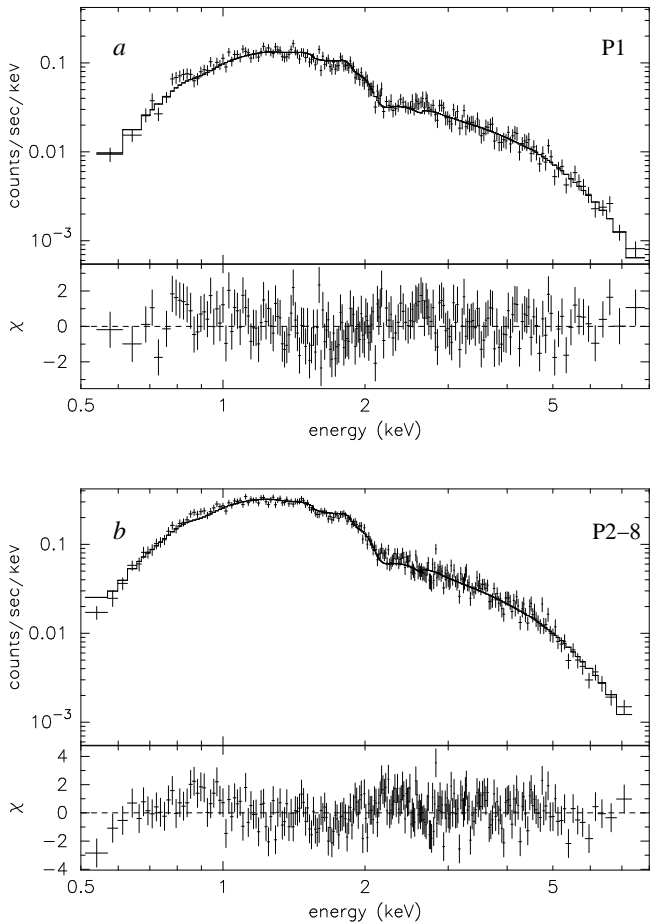,height=4.8in,angle=270, clip=}
\hfil\hfil}}
\caption{ACIS-S spectra in the Regions P1 (a) and P2-P8 (b), together
with  power-law fits.}
\label{f:spec_pp}
\end{figure}

\begin{figure}[tbh!] 
\centerline{ {\hfil\hfil
\psfig{figure=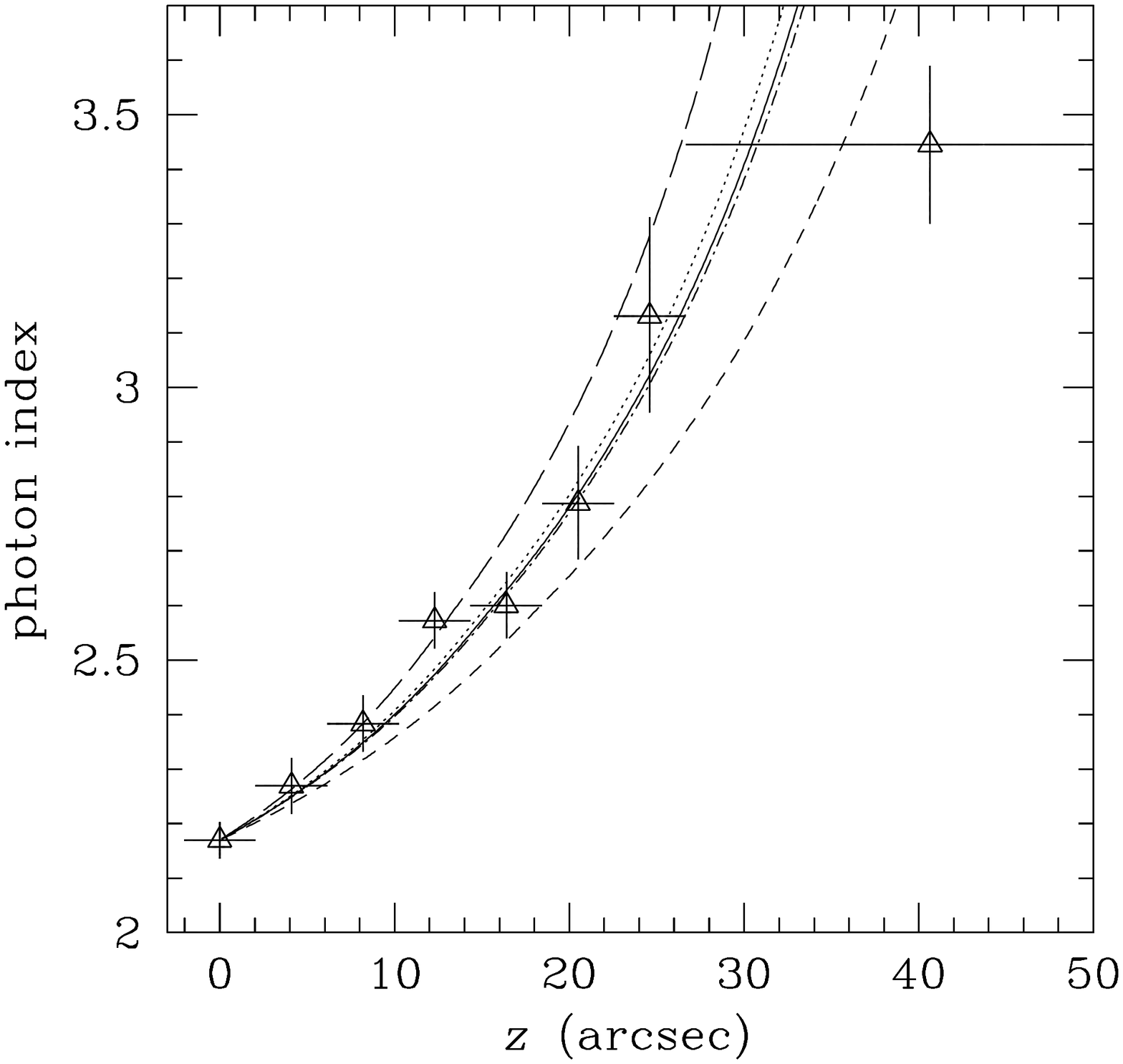,height=3in,angle=0, clip=}
\hfil\hfil}}
\caption{Variation of the power-law photon index of the nonthermal
tail as a function of  the distance from the pulsar.
Curves are plotted according to the particle ejection model
represented by Eq.~(\ref{eq:index}) with various $u$ and $\Bp$
parameter combinations:
$0.5c$ and $0.9\E{-4}\G$ (dotted curve),
$0.6c$ and $0.9\E{-4}\G$ (short-dashed),
$0.6c$ and $1.0\E{-4}\G$ (solid),
$0.6c$ and $1.1\E{-4}\G$ (long-dashed),
and $0.7c$ and $1.1\E{-4}\G$ (dot-dashed).
}
\label{f:spec_index}
\end{figure}

We examine the spectral properties of the key nonthermal features, 
individually. Fig.~\ref{f:spec_p0} shows the net spectrum of the halo, 
extracted from Region P0 (a local background
described in \S~\ref{sec:spec_small} is subtracted).
In addition to a dominant contribution from a power law with a photon index of 
$\Gamma\sim 1.7-2.1$,
there is clearly a thermal component, as evidenced
by the presence of the $\sim1.34\keV$ Mg~XI~K$\alpha$ line.
Characterized by a thermal plasma  ($kT\sim0.6$--$1.2\keV$), this component
accounts for about 20\% of the total emission from Region P0.
The exact contribution from this thermal 
component is uncertain, somewhat depending on the background 
subtraction. The component, rich in magnesium ($\sim0.6$--4.5 solar),
may represent an enhanced surrounding thermal plasma emission
projected along the line of sight.

Fig.~\ref{f:spec_pp} shows the net spectra
of regions P1 (with the pulsar excluded) and P2--P8 (Fig.~\ref{f:reg}). The
P1 spectrum is dominated by the SW-NE oriented bar 
($\sim 2.4''\times6.8''$), which has a very 
high-surface brightness. We can thus 
divide the bar region into individual segments, but find no 
statistically significant spectral 
variation on either side of the pulsar. 
The spectrum of Region P1 can be well fitted
($\chi^2/{\rm d.o.f.} =206.3/191$) by a power-law with a photon index of 
$2.18^{+0.59}_{-0.53}$
and a foreground absorption $\NH\sim6.1^{+2.7}_{-2.4}\E{21}\cm^{-2}$.
The accumulated spectrum of the P2--P8 regions may also be characterized
approximately ($\chi^2/{\rm d.o.f.} =318.1/241$) by a power-law 
with an index of $2.62\pm0.04$
and a hydrogen column $\NH\sim (6.3\pm0.2)\E{21}\cm^{-2}$.
The $\NH$ values are consistent with those inferred from the spectral 
analyses of the entire remnant (\S~\ref{sec:glo}) and the pulsar 
(\S~\ref{sec:psr}).
The spectra from the individual P1--P8 regions, fitted well with
the power law, show a systematic change 
of the photon 
index along the major axis of the nebula (Fig.~\ref{f:spec_index}).
Here we have fixed $\NH$ to the value $6.1 \E{21}\cm^{-2}$;
if  $\NH$ is allowed to vary in the fit, 
the result is very similar, albeit with larger error bars.

\begin{center}
\begin{deluxetable*}{lcccc} 
\tabletypesize{\footnotesize}
\tablecaption{Spectral Fits for Individual Thermal Features}
\smallskip
\tablewidth{0pt}
\tablehead{
\colhead{parameter} &
  \colhead{L1--3} & \colhead{D1} & \colhead{D2} & \colhead{D3} 
}
\startdata
net count rate ($10^{-2}$ counts$\ps$) & $1.79\pm0.10$ & $4.06\pm0.11$
 & $3.70\pm0.10$ & $3.76\pm0.13$ \\
$\chi^2$/d.o.f. & 23.6/24 & 84.3/60 & 87.1/62 & 45.4/40 \\
$\NH$ ($10^{21}\cm^{-2}$) & $3.8\pm0.6$ & $7.4^{+1.0}_{-0.4}$
 & $4.8^{+0.8}_{-0.7}$ & $9.3^{+1.6}_{-2.8}$ \\
$kT$ (keV) & $1.04^{+0.89}_{-0.31}$ & $1.47^{+0.59}_{-0.40}$
 & $1.22^{+0.41}_{-0.31}$ & $0.23^{+0.15}_{-0.07}$ \\
{[O/H]} & \nodata & \nodata & \nodata
  & $0.79^{+1.29}_{-0.40}$ \\
{[Ne/H]} & \nodata & $1.38^{+0.36}_{-0.30}$ & $0.96\pm0.26$
  & $0.51^{+0.50}_{-0.20}$ \\
{[Mg/H]} & \nodata & $2.23^{+1.10}_{-0.84}$ & $0.50^{+0.31}_{-0.15}$
  & ...  \\
$n_e t_i$ ($10^{10}\cm^{-3}\s$) & $5.33^{+8.30}_{-3.47}$ & $0.68^{+0.45}_{-0.15}$
  & $2.39^{+0.93}_{-0.96}$ & $2.30^{+8.13}_{-1.38}$ \\
$fn_e\nH V$ ($10^{57}\cm^{-3}$)
  & $3.20^{+2.05}_{-1.15}$ & $7.33^{+4.42}_{-2.50}$ & $4.76^{+2.50}_{-1.44}$
  & $540 (<3460)$\\
flux ($10^{-13}\ergs\cm^{-2}\ps$) & 2.0 & 14.5 & 6.0
  & 313\\
$\nH f^{1/2}$ ($\cm^{-3}$)$^{\rm a}$ & 0.22 & 0.56 & 0.67 & 3.8
\enddata
\tablecomments{The fluxes are in the 0.5--$10\keV$ band.}
 \tablenotetext{a}{\phantom{0} In the estimate of the densities, we assume
 cuboidal volumes for rectangular regions L1--L3, with sizes
 $8.0\times7.5\times12\parsec^3$, $14\times5.2\times12\parsec^3$,
 and $7.3\times2.9\times12\parsec^3$, respectively, 
 and oblate spheroids for elliptical regions D1--D3, with 
 half axes $8.6\times8.6\times2.2\parsec^3$, $5.9\times5.9\times2.1\parsec^3$,
 and $7.2\times7.2\times5.0\parsec^3$, respectively.}
\label{t:small}
\end{deluxetable*}
\end{center}

\subsection{Diffuse Emission: Thermal Component}\label{sec:spec_small}

\begin{figure*}[tbh!] 
\centerline{ {\hfil\hfil
\psfig{figure=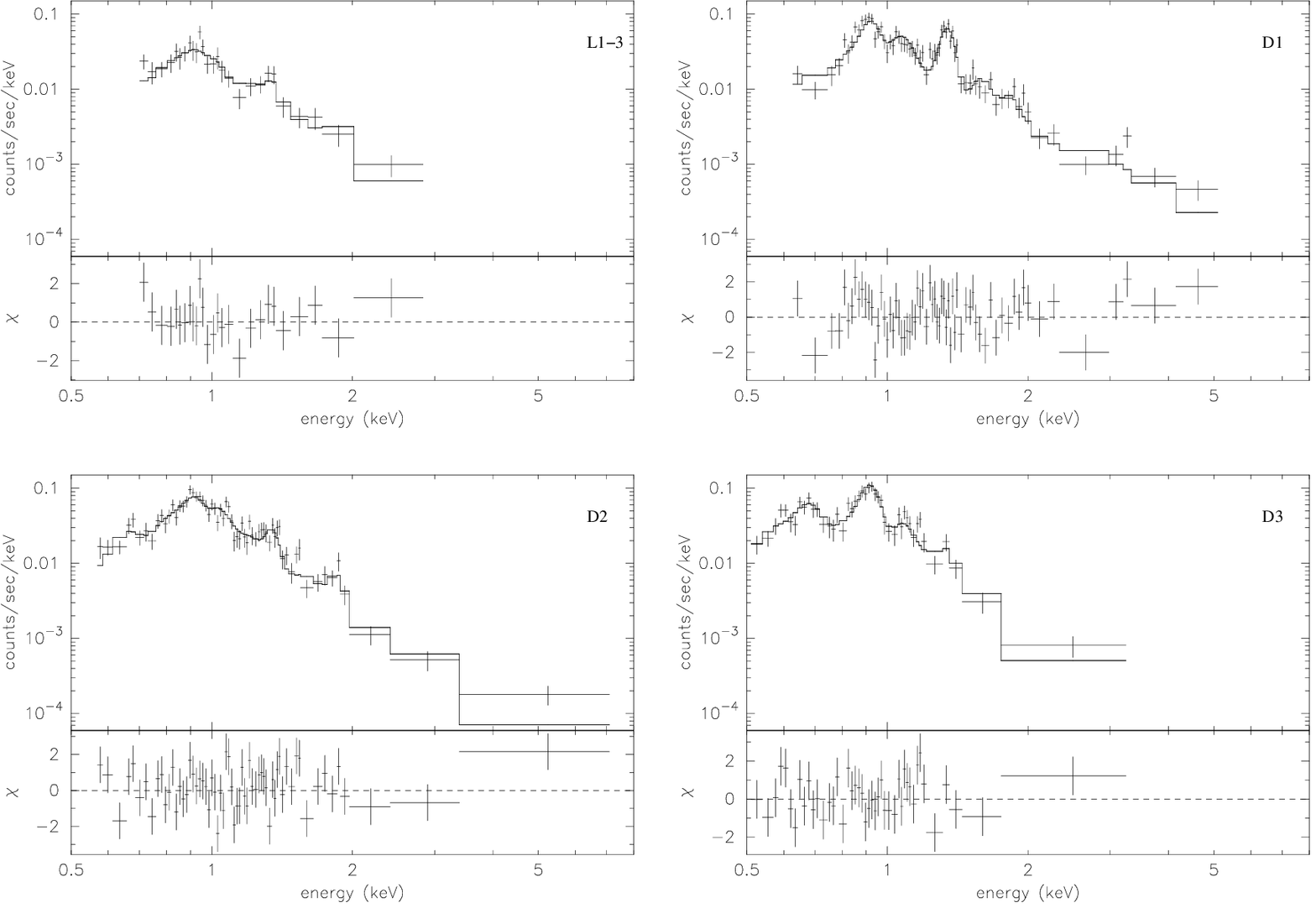,height=4.8in,angle=270, clip=}
\hfil\hfil}}
\caption{ACIS-S spectra of selected thermal substructures (see 
Fig.~\ref{f:reg}: L1--3, D1, D2, and D3). All these
spectra are fitted with the NEI model (Table~\ref{t:small})
and are re-binned to achieve a background-subtracted
S/N ratio of 3.}
\label{f:spec_th}
\end{figure*}

We characterize the properties of the diffuse X-ray emission away from 
the nonthermal cometary nebula by analyzing the ACIS-S spectra of 
individual regions of \snr\ (Fig.~\ref{f:reg}; Fig.~\ref{f:spec_th}). 
The spectrum of Regions L1--3, excluding the background estimated in 
Regions B1 and B2, is used to characterize the general low-surface
brightness emission from the SNR.
For the spectral analysis of the substructures (D1, D2, and D3),
we adopt the L1--3 spectrum as the local background.
All these spectra show distinct emission lines,
indicating predominant contributions from optically-thin 
thermal plasma. Indeed, the spectra can all be fitted well with 
the NEI model. Similar to the analysis of the accumulated
spectrum from the entire SNR (\S~\ref{sec:glo}), the model fits 
allow the abundances of key line elements to vary, while leaving
the rest of the metal abundances fixed to the LMC value (0.3 solar). 
Table~\ref{t:small} summarizes the net count rates and results of 
the spectral fits, as well as the inferred energy fluxes and 
hydrogen number densities of the individual regions. 
The densities are derived on the assumption that the 3-D shapes 
of the elliptical regions (D1-D3) are oblate ellipsoids
and those of the rectangular regions (L1--L3) are cuboids
with the average line-of-sight size to be the remnant radius.
The deviation from this assumption and the non-uniformity of 
the X-ray-emitting plasma are consolidated into the ``filling 
factor,'' $f$, of the individual regions.
The regions show significant spectral variation among them. 
While the metal abundances in Regions L1--3 are consistent with the 
LMC value, the substructures all exhibit significant metal enrichment
and variations in physical conditions.
For example, Region D3 shows both a strong O~VIII K$\alpha$ line 
and a low plasma temperature, which explains the softness of its
emission (e.g., Fig.~\ref{f:im_color}b).
Regions D1 and D2 show comparable temperatures, densities, and 
neon abundances, but Region D1 is also particularly abundant in Mg,
seen in both the spectrum and the EW map, and its ionization time-scale
is significantly smaller than that of Region D2.

It is possible that part of the diffuse X-ray emission may have origins 
other than the N157B SNR.
However, there is little evidence for a spatial correlation of the diffuse
X-ray intensity with stellar concentrations (e.g., Fig.~\ref{f:im_color}c), 
indicating that the direct stellar emission and the gas heating due to 
stellar winds from massive stars (e.g., wind-wind and wind-cloud collisions) 
do not contribute significantly. The X-ray emission from the 
pre-existing superbubble is substantially softer than what is inferred from 
the SNR and has a very low-surface brightness (Smith \& Wang 2004). Our
analysis is based on the enhanced emission above the local background, which
includes the superbubble contribution.
 
\section{Discussion}

Based on our analysis, we now attempt to present a coherent 
interpretation of the characteristics of \snr: the exceptionally 
large sizes of both the thermal and nonthermal components and the lack of 
a well-defined outer shell as well as the
distinct spatial/spectral substructure and metal enrichment pattern. 

\subsection{Environmental effect}

\begin{figure}[tbh!] 
\centerline{ {\hfil\hfil
\psfig{figure=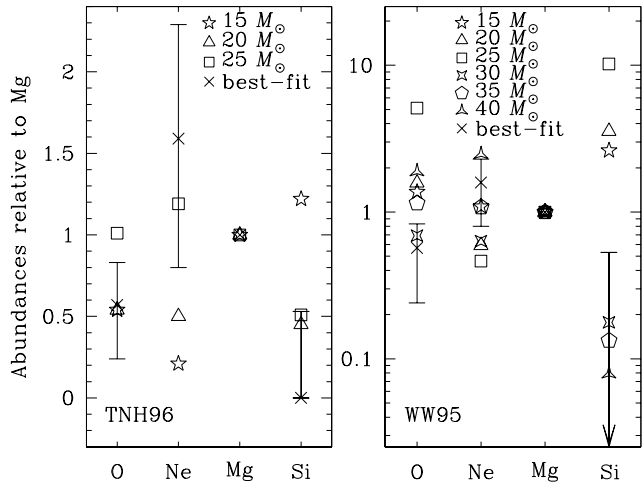,height=2.75in,angle=270, clip=}
\hfil\hfil}}
\caption{The best-fit abundances relative to Mg,
compared with the predictions from core-collapsed SN models 
(TNH96 and WW95).
Note that because the best-fit Si value is only $\sim10^{-8}$,
it is invisible in the right panel (with a logarithmic scale).
}
\label{f:abund}
\end{figure}

\snr\ is presumably associated with the OB association LH99, as they have
comparable foreground absorption/extinction (WG98), in addition to
their apparent superposition in the sky. 
While the characteristics of \snr\ appear to be a manifestation of
its association with LH99, the new X-ray results provide new insights into 
the nature of the SN progenitor and the properties of the ambient
environment.  First, we can use the observed pattern of 
metal enrichment to constrain the mass of the progenitor star. 
As the presence of a pulsar implies a core-collapse nature of 
the SN, we plot in Fig.~\ref{f:abund} the relative metal 
abundances predicted from such SN models for stars with low
metallicities (Thielemann, Nomoto, \& Hashimoto 1996, hereafter 
TNH96; Woosley \& Weaver 1995, hereafter WW95).
The enrichment is compatible with the TNH96 model for a 
progenitor mass of 20--25$\Msun$, or the WW95 models for
a progenitor mass of 30--35$\Msun$, $Z=0.1Z_{\odot}$,
and an explosion energy of $\sim 1.2\E{51}\ergs$.
Although these two model predictions do not overlap, they
both indicate a progenitor mass of $\gsim20\Msun$.
Recent models taking into account stellar mass loss suggest
that neutron stars can be produced by metal-poor progenitors
only if the progenitor mass is $<25\Msun$ (Heger et al.\ 2003). 
This mass limit, while consistent with the above estimate from 
the TNH96 model, is lower than the masses of the O3 and O5 stars
that still exist in LH99 (Schild \& Testor 1992). The earlier explosion 
of a less massive star thus implies that the star formation in
the vicinity of LH99 is not coeval.  Further evidence for a 
prolonged star formation has recently been provided by the 
massive proto-stars revealed in {\sl Spitzer} observations; for 
example, the brightest point source in the {\sl Spitzer} 8 $\mu$m 
image is a proto-star (Fig.~\ref{f:im_color}a).

We next examine how the co-existence with the OB association LH99
may have influenced the evolution of \snr. Fig.~\ref{f:im_color}a 
shows a {\sl Spitzer} IRAC 8 $\mu$m view of the large-scale environment
of \snr.  The 8 $\mu$m band image is dominated by emission from 
polycyclic aromatic hydrocarbons (PAHs).  As PAHs can be easily
dissociated by UV radiation from O stars, the 8 $\mu$m emission
traces \ion{H}{1} gas just outside the ionization front.
The 8 $\mu$m image shows that \snr\ is inside a large cavity that
might be a superbubble produced by the mechanical energy input from 
the OB association LH99.  
The existence of a blister-type superbubble around \snr\ has
also been suggested previously based on the velocity 
structure of dense ionized gas (Chu et al.\ 1992).
This IR cavity might also be connected with the cavities and ``tunnels''
powered by multiple OB associations, including also LH90 to the east,
as indicated by the presence of a large-scale ($\sim 20^\prime$) 
low-surface brightness X-ray emission in the region (Smith \& Wang 2004).
The exact shape of the mid-infrared cavity, let alone its boundaries, is
very uncertain; there may be substantial projection effect.
In morphology, the pulsar, presumably close to the explosion center,
is located at
the west edge of the cavity. The large-scale diffuse 
X-ray emission extends into the cavity. 
The expansion of the SNR into such a low-density hot medium at least 
partly explains its unusual large size and its lack of a well-defined
outer shell.  The hot ambient medium has a high isothermal sound 
velocity, thus the shocks are expected to be weak with small Mach
numbers, and the blastwave may become thermalized and deposit the
bulk of the SNR energy into the thermal energy of the hot medium
(Tang \& Wang 2005). 
The relatively low pressure inside the SNR further allows for 
the expansion of the PWN to an exceptionally large volume. 

While the overall morphology of N157B is apparently affected by 
the large-scale superbubble environment, 
the presence of a dense cloud close to the explosion site 
seems to play a critical role in shaping the dynamics of the materials
around the pulsar. This provides an alternative explanation for
the trailing PWN structure, which we previously proposed to be
due to pulsar motion (WG98; Wang et al.\ 2001).
Probably, the trailing tail is not necessarily
produced by the passage of the moving pulsar; and an existing example
may be the Duck X-ray Nebula, where the tail is ejection of pulsar
wind particles at weak relativistic bulk velocity confined by 
the ambient medium ram-pressure (Kaspi et al.\ 2001).
In SNR~N157B, the pulsar appears to be near the explosion center
and have not moved out of the SNR, at least in projection.
The size of the nonthermal nebula, especially its head, is large, which is
somewhat difficult to explain in a supersonic relative motion scenario,
although the physics of the shocked pulsar wind is quite uncertain,
complicated by its non-spherical symmetry. Instead, the subsonic
ram-pressure confinement can be consistent with the gas motion
relative to the pulsar (see below and \S\ref{sec:ther_nat}).
The cloud is partly seen as the strongest diffuse 
mid-infrared emission peak south to the pulsar in Fig.~\ref{f:im_color}a.
The cometary PWN tail points to the northwest,
almost parallel to the long dimension of the cloud or the southwest edge
of the superwind cavity (Fig.1a),
consistent with the relatively narrow extent of the thermal
emission on the southwest side of the PWN (Fig.1b and Fig.1d).
The northern part of the cloud, not apparent in Fig.~\ref{f:im_color}a, 
is bright in optical (Fig.~\ref{f:im_color}c). The intensity of the
enhanced diffuse optical light shows a
reasonably good morphological similarity to the relatively high 
surface-brightness X-ray emission in the 0.3--1.5 keV band. The dust
lane at the southeast side of Fig.~\ref{f:im_color}c casts a shadow
in the emission distribution and corresponds to harder X-ray emission
(Fig.~\ref{f:im_color}b), apparently due to the X-ray absorption
(but the limited counting statistics and the uncertainty in the
background subtraction make it hard to quantify the absorbing gas column).
In fact, the association of \snr\ with this optical
nebula has already been suggested by Chu  et al.\ (1992), based on its
kinematics. In view of these observational factors,
we suggest a physical association that is produced by 
the impact of \snr\ on the cloud. Although the shape and boundary
of the mid-infrared cavity are very uncertain (considering substantial
projection effect) and the exact interplay (the 
structure and dynamics) between the two cannot yet be determined,
this impact likely led to a strong reflection shock, which
has probably swept through and been further reflected around much of the PWN. 
It is also possible that there is a reflection from the east edge
of the cavity, although we find little optical evidence for a 
strong interaction. The resultant heating by reflection shock and possibly
large bulk flow of the SNR gas may thus be responsible for both the 
observed one-sided morphology of the PWN and the thermal 
emission enhancement (like a sheath) around the southern part of the PWN. 
The enhancement east of the pulsar (Region D2) is particularly strong just
($\sim 10^{\prime\prime}-20^{\prime\prime}$) east of the pulsar
and is bordered by a dust lane farther to the east.
It shows no evidence for nonthermal emission. This part of the
enhancement is clearly edge-enhanced and has corresponding optical
brightening (Fig.~\ref{f:im_color}c), probably representing
the strongest ongoing reflection of the SNR from the cloud.
The high-pressure of the reflected gas may have greatly influenced
the confinement of the pulsar wind and hence the direction
of its outflow toward the northwest. This interpretation of the PWN
and its surrounding medium is similar to the scenarios proposed
by WG98 and van der Swaluw et al.\ (2003): the observed one-sided
morphology is a result of an asymmetric ram-pressure confinement. 
But, the present interpretation of the strongly one-sided PWN
morphology does not require a fast proper motion of the pulsar;
the ram-pressure (or the relative motion between the pulsar and 
the surrounding medium) can naturally arises from the highly
asymmetric reflection shock of the SNR (see more discussion below).
Having proposed this alternative, we do not abandon the pulsar
proper motion scenario, which is still plausible.

\subsection{Evolutionary Stage of SNR \snr} \label{sec:ther_nat}

The properties of \snr\ is determined not only by its environment
but also by its specific evolutionary stage, although its age is still
very uncertain. The spin-down age of the associated pulsar \psr\ is 
$\sim5\E{3}\yr$, as derived from the pulse period and its derivative,
assuming a negligible initial pulsar period and a dipole magnetic
field configuration (Marshall et al.\ 1998).
These assumptions may not be valid for young pulsars in general and 
for this pulsar in particular because it displays repeated episodes 
of strong timing glitches (Marshall et al.\ 2004).  
Nevertheless, the spin-down age is roughly consistent with the large
size of the remnant (12 pc in radius for a total angular extent of
$\sim100''$).
From the ionization timescale ($n_e t_i\sim2.4\E{10}\cm^{-3}\s$;
Table~\ref{T:ent}) and 
the mean gas density ($\nH\sim0.45f^{-1/2}\ru^{-1/2}\cm^{-3}$ where 
$\ru=R/12\parsec$, assuming $\nel\sim1.2\nH$), 
we may estimate an average ionization age of the X-ray-emitting
plasma in the remnant (i.e., the time elapsed since it was shocked)
as $t_i\sim2\E{3} f^{1/2}\ru^{1/2}~\yr$, 
which is considerably shorter than the pulsar spin-down age, especially 
because $f$ can be small.  This discrepancy may be understood if the
X-ray emission originates primarily from the plasma that is recently
shock-heated.

Indeed, the thermal properties of \snr, as inferred from
\S~2.4, suggest that the X-ray-emitting gas consists of various 
metal-enriched clumps embedded in a more diffuse plasma. 
This diffuse plasma is distributed more widely than the clumps and
also seems to have both a nominal
LMC metal abundance and a larger average ionization age, which is
comparable to the spin-down age of the pulsar. Therefore, the plasma
may mostly represent the shocked ISM, whereas the clumps are mainly due
to SN ejecta. Unfortunately, the quality of the data 
is not adequate to make a quantitative decomposition of these two
components. We could only make a crude estimate of
the thermal energy as $\sim3\E{50}f^{1/2}\ru^{3/2}\ergs$ and the mass 
as $\sim110f^{1/2}\ru^{3/2}\Msun$. This estimate is subject to the
uncertainty in the effective filling factor, which cannot be determined
and is small, probably on the order of $\lsim0.1$. This gas mass is comparable
to several solar masses expected for the SN ejecta and is thus consistent 
with the significant metal-enrichment detected. Depending on
the $f$ value, the thermal energy may 
be considerably smaller than the canonical SN energy 
($\sim 10^{51}$ ergs). A substantial fraction of the energy may still be
in the kinetic form, carried primarily by a weak outward blastwave into
the low-density hot superbubble interior (\S~3.1; Tang \& Wang 2005). 

Among the substructures considered, the enhancement in Region D3 
shows distinct characteristics: a partially round shape with 
a size of $\sim30''$ or $7\parsec$,
a distinctly low gas temperature ($\sim0.2\keV$), 
a high density ($\sim4f^{-1/2}\cm^{-3}$), a
low ionization age ($\sim160f^{1/2}\yr$), and an
apparent overabundance in oxygen. This feature may represent 
a clump of oxygen-rich ejecta recently heated by a low-velocity 
reverse shock ($\sim400\km\ps$), as inferred from the gas temperature.  
However, this hypothesis cannot explain the morphology or size,
as the ionization time scale (Table~4) would suggest a velocity
of $\sim 4 \times 10^4 f^{-1/2}{\rm~km~s^{-1}}$,
much higher than the value inferred from the plasma temperature.
Of course, the NEI plasma we used may be an over-simplification
for the modeling of the spectrum (Fig.~\ref{f:spec_th}),
and the inferred ionization time scale may be misleading. Without 
the constraint on the time scale, the D3 enhancement may simply be part of
the shocked SN ejecta, possibly ballooning out from the high-pressure SNR 
interior into the superbubble. In this case, the relatively low temperature 
can be naturally explained by the adiabatic cooling of the expanding gas.
Alternatively, D3 may arise from a separate oxygen-rich SN, which 
is quite possible with the presence of the OB association LH99. 
In this case, the partially round morphology can naturally be explained as a 
coherent shell of reverse-shocked SN ejecta. The age of the remnant
should be comparable to the ionization age inferred above.
The reverse shock velocity is expected to be substantially lower 
than that of the forward blastwave, which is probably very weak in 
X-ray emission.  If the remnant is also expanding inside the superbubble, 
the lack of an apparent radio or optical counterpart is then expected.
The absence of a PWN associated with the remnant further suggests that 
the stellar progenitor of the SN is more massive ($\gtrsim 25 M_\odot$; 
Heger et al.\ 2003) and that the stellar remnant may be a black hole.
In fact, we are also a bit uncomfortable about involving a second SNR
in the interpretation. But if we accept the spectral results (mainly the
small ionization parameter) from the NEI model fit to the D3 spectrum,
we find no better solution than a new SNR.

As discussed in \S~3.1, a ``sheath'' of the thermal X-ray enhancement
(including D1+D2) around the PWN is likely a result of its interaction 
with the material reflected off the dense cloud. 
The physical processes involved are similar to those in the
interaction between an SNR reverse shock and a PWN of a fast moving 
pulsar, as simulated by van der Swaluw et al.\ (2004). 
Before such an interaction, the PWN consists 
of two main components, the shocked pulsar wind material and the 
swept-up SN ejecta between the forward shock and the contact 
discontinuity (Reynolds \& Chevalier 1984).
The thermal emission from the swept-up SN ejecta is expected to be
weak and is confined within a relatively thin shell.  The passage
of the strong reflection shock (e.g., stage 2 in van der Swaluw et al.\
2004) compresses the PWN, sweeps part of it into the trailing tail,
and is meanwhile further reflected forward.
Consequently, both the previous undisturbed and the swept-up 
SN ejecta are compressed and heated at the 
front side of the PWN. The density and temperature
in this ``sheath'' around the PWN should be enhanced 
by a factor of 1--2.5 (compared to those in the upstream 
gas), depending on the Mach number of the reflection shock relative 
to the pulsar (Spitzer 1982; Hester \& Cox 1986).
Such density and temperature contrasts are consistent with our 
measurements, albeit with large uncertainties (Table~\ref{t:small}).
We thus conclude that the interaction between the reflected material
and the PWN provides a natural interpretation of both its strongly
one-sided morphology and its surrounding thermal 
X-ray emission enhancement.

\subsection{Physical Properties of the PWN} \label{sec:nonther_nat}

Based on the spectroscopic results of various nonthermal components 
of \snr, we attempt to determine their nature and quantify their physical 
parameters. The bar around the pulsar \psr\ has been proposed to be a 
torus (Wang et al.\ 2001), similar to those detected around 
several well-known young pulsars [e.g., Crab (Weisskopf et al. 2000),
Vela (Helfand, Gotthelf, \& Halpern 2001), PSR 0540-69
(Gotthelf \& Wang 2000), and G54.1+0.3 (Lu et al.\ 2002)]
and believed to represent the termination shocks of the 
toroidal pulsar winds (e.g., Ng \& Romani 2004). This
proposal is consistent with the results
from our spectral analysis of the bar. 
The lack of a systematic spectral variation across
the bar simply reflects similar particle acceleration 
and synchrotron emission throughout the termination shock region,
just as in the Crab nebula  (Mori et al.\ 2004). 
The power-law photon index
inferred from the analysis also agrees well with the values
($\sim 2$) observed in the other toruses (Gotthelf 2003). The centrally
enhanced bar-like morphology can be naturally 
explained by the edge-on orientation of the torus because
of the expected Doppler boosting, which
depends on the angle between the bulk motion velocity of 
shocked pulsar particles and our sight line (e.g., 
Komissarov \& Lyubarsky 2003). Similar effects have been observed in 
both the Crab nebula and G54.1+0.3 (Lu et al.\ 2002).  
The size of the torus should be comparable to the radius of
the termination shock (modified from Chevalier 2000),
\begin{equation}
r_t=\lt(\frac{6\varepsilon_B\dot{E}}{c\sin\theta_{\rm t}}\rt)^{1/2}B^{-1},
\end{equation}
where $\varepsilon_B$ denotes the magnetic field fraction of the pulsar 
wind luminosity ($\sim0.5$ in the equipartition case)
and $\theta_{\rm t}$ is the half angular span of the
toroidal wind (i.e., $4 \pi \sin\theta_{\rm t}$ is the solid angle
subtended at the pulsar by the torus).
The pulsar spin-down power $\dot E = 4.8 \times 10^{38}$~ergs~s$^{-1}$
(Wang et al.\ 2001; Mignani et al.\ 2005) is adopted.
The magnetic field $B$ just behind the shock is not well understood. But if
it is comparable to the average strength ($\sim 2 \times 10^{-4}$ G)
in the PWN and $\theta_{\rm t}=20^{\circ}$ (as estimated in Wang
et al.\ 2001), we can then estimate $r_t\approx 0.6\parsec$ or $2\farcs4$.
This estimate is comparable to the observed overall dimension
($\sim2\farcs4\times6\farcs8$) of the bar, and thus both the 
morphological and spectral characteristics of the bar are 
consistent with its being an edge-on torus resulting from the 
terminal shock of the pulsar toroidal wind.

The \snr\ PWN as a whole shows a featureless and hard spectrum. 
The spectral steepening along the major axis of the tail
indicates significant synchrotron cooling.
Similar spectral properties have been seen in other SNRs,
such as G21.5$-$0.9 (Slane \etal.\ 2000; Warwick et al.\ 2001,
Safi-Harb et al.\ 2001), 3C58 (Torii et al.\ 2000), IC443 
(Bocchino \& Bykov 2001), and possibly in W44 (Petre, Kuntz, 
\& Shelton 2002); however, their X-ray softening has not been 
quantitatively compared with models.
The bright and long trailing pattern of the \snr\ nebula makes 
it possible to quantitatively compare observations with models.

We construct a simple 1-D toy model,
similar to the spherically symmetrical treatment by Amato et al.\ (2000).
In the 1-D approximation of the \snr\ PWN, 
the shocked wind particles stream off the bar
along the downstream tail at a  
speed $u$ within a cylinder of radius $r$. The differential 
particle density distribution as a function of the distance from the 
injection point (the bar) $z$ is then 
\begin{equation}
N(E,z)=\frac{J(E_{\rm ini})}{\pi r^2}\lt|\parder{t_{\rm ini}}{z}\rt|\,
 \lt|\parder{E_{\rm ini}}{E}\rt|,
\end{equation}
where $E_{\rm ini}$ and $t_{\rm ini}$ are the initial energy and
time of the injected particle, while the injection rate $J(E_{\rm ini})$
is assumed to be a power law,
\begin{equation}
J(E_{\rm ini})=KE_{\rm ini}^{-p}.
\end{equation}
If the bulk velocity $u$ is assumed to be constant (i.e., $z/t$), 
the energy loss of the particles resulting from the synchrotron radiation
is then 
\begin{equation}
\totder{E}{t}=-c_1\Bp^2E^2,
\end{equation}
where $c_1=4e^4/(9m_e^4c^7)$
and $\Bp$ is the magnetic field component
perpendicular to the particle motion.
Integration of the above equation gives
$E_{\rm ini}=E/(1-E/E_{\rm m})$, where
\begin{equation}
E_{\rm m}=u/(c_1\Bp^2z)
\end{equation}
marking the maximum energy of the particles after the synchrotron loss
at $z$.
Then the particle density distribution is given by
\begin{equation}
N(E,z)=\frac{K}{\pi r^2 u}E^{-p}\lt(1-\frac{E}{E_{\rm m}}\rt)^{p-2}.
\end{equation}
The synchrotron emissivity of the particles is
$j_{\epsilon}=(c_1\Bp^2E^2)N(E,z)dE/d\epsilon$, where the photon energy is
related to the particle energy as $\epsilon=c_2h\Bp E^2$
[where $c_2=0.29\times3e/(4\pi m_e^3c^5)$].
Therefore
\begin{equation}
j_{\epsilon}=\frac{c_1}{2c_2h}\frac{K}{\pi r^2 u}\Bp E^{1-p}
        \lt(1-\frac{E}{E_{\rm m}}\rt)^{p-2}.
\end{equation}
The dependence of the emission's photon index,
\begin{equation}
\Gamma\equiv -(1+d\ln j_{\epsilon}/d\ln\epsilon),
\end{equation}
implicitly on $\epsilon$ and $z$ (via $E$ and $E_{\rm m}$, respectively)
is derived as
\begin{equation}
\Gamma=\frac{p+1}{2}+\frac{p-2}{2}\frac{E/E_{\rm m}}{1-E/E_{\rm m}}
\end{equation}
The above equation shows that the spectral index varies with both
the photon energy and the distance from injection point.
Since the spectral indices are determined in a broad energy range,
to compare with the observed values, we calculate a weighted spectral
index,
$\bar{\Gamma}=\int_{\epl}^{\min(\epu,\epm)}$ $[j(\nu,z)/\epsilon]\Gamma(\nu,z)
  \,d\epsilon / \int[j(\nu,z)/\epsilon]\,d\epsilon$, namely
\begin{equation}\label{eq:index}
\bar{\Gamma}=\frac{p+1}{2}+\frac{p-1}{2}
 \frac{\lt(\sqrt{\epm/\epl}-1\rt)^{p-2}-\lt(\sqrt{\epm/\min(\epu,\epm)}-1\rt)^{p-2}}
     {\lt(\sqrt{\epm/\epl}-1\rt)^{p-1}-\lt(\sqrt{\epm/\min(\epu,\epm)}-1\rt)^{p-1}},
\end{equation}
where $\epl$ and $\epu$ are the lower and upper energy limits of the 
spectrum used, while $\epm$ is the photon energy corresponding to $E_{\rm m}$,
\begin{equation}\label{eq:eb}
\epm=\lt(\frac{z}{17\parsec}\rt)^{-2}\lt(\frac{u}{c}\rt)^{2}\lt(\frac{\Bp}{10^{-4}\,{\rm G}}\rt)^{-3}\keV.
\end{equation}

Eq.~\ref{eq:index} can be compared with the measured values.
We fix the photon index at the injection point (P1), $\Gamma = 2.17$
(for $\NH=6.1\E{22}\cm^{-2}$),
which corresponds to $p =3.34$, and adopt $\epl=0.5\keV$ and $\epu=7\keV$.
Fig.~\ref{f:spec_index} shows predicted curves, assuming the 
magnetic field to be around the mean PWN value (Wang et al. 2001).
These curves match the measured values well, suggesting $u\gsim0.5c$,
comparable to the sound speed of an ultra-relativistically 
hot plasma ($c_s = c/\sqrt{3}$).  Similar high downstream flow speeds  
are observed or inferred at the terminal shocks of the Crab nebula 
(Hester et al.\ 2002) and G54.1+0.3 (Lu et al.\ 2002); but in these 
cases, the flows are presumably quickly damped and energy is then transported
primarily due to particle diffusion. In an one-sided pressure 
confined PWN, the downstream flow is re-directed into a bulk motion
at a similar or higher speed, as demonstrated analytically (e.g., Wang,
 Li, \& Begelman 1993) and in simulations (e.g., Bucciantini et al. 2005).
Therefore, a strong pressure-confinement of the PWN is needed
to give a consistent interpretation of both the exceptionally large
linear size and the observed spectral steepening of the nonthermal
component of the \snr\ PWN.

\section{Summary}

Based on our on-axis {\sl Chandra} ACIS observation, we have conducted
a detailed spatially-resolved X-ray spectroscopy of \snr.
Combining the X-ray analysis with the optical and mid-infrared
observations, we have explored the physical origins of various
distinct characteristics and their implications.
Our main results and conclusions are summarized as follows:

\begin{enumerate}

\item We detect eight point-like X-ray sources in the ACIS observation. Two
of these sources in the vicinity of the pulsar  \psr\
have near-IR and optical counterparts, which are identified
as massive stars in the LMC. The relatively high X-ray luminosities of
the two sources indicate that they are probably colliding stellar wind
binaries. A comparison of the X-ray and 2MASS positions of the two sources
suggests that the
absolute astrometry of the ACIS observation, hence the X-ray centroid
position of the pulsar, is better than about 0\farcs5.

\item We confirm the nonthermal nature of the various high-surface brightness 
X-ray features, such as the bar, tail, and halo, as suggested in the
previous morphological and integrated spectral studies.
The bar shows no significant spectral shape variation and
most likely represents the terminal shock of a toroidal pulsar wind.
The strong spectral steepening observed in the tail is consistent with the
synchrotron radiation loss and suggests a bulk velocity
of outflowing pulsar wind particles to be $\gsim0.5c$.

\item The thermal emission is unambiguously resolved to arise from a
region much larger than the nonthermal PWN and accounts 
for about one-third of the total X-ray emission in the 0.5-10 keV band.
The thermal emission consists of various clumps embedded in 
a diffuse X-ray-emitting plasma.
The clumps are over-abundant in $\alpha$ elements and thus represent 
the SN ejecta.
The over-abundance pattern suggests a mass $\gsim20\Msun$ for the 
SN progenitor.
However, a partially-round-shaped soft X-ray clump in the outer region of
\snr\ shows distinct spectral
characteristics, unusually low in both temperature and ionization 
time scale and over-abundant in oxygen. This clump may thus 
represent a separate young SNR.

\item The thermal emission of \snr\ is enhanced around the nonthermal PWN,
particularly near its head.
This enhancement is perhaps a result of the recent compression and
heating of the ejecta due to the SNR reflection
from a nearby dense cloud. This interaction of the SNR reflection and
the PWN naturally explains its one-sided morphology.

\item The relatively weak thermal emission,
the lack of a rim-brightened outer blastwave,
and the large sizes of both the thermal and nonthermal components
all point to a scenario that SNR \snr\ is
expanding into a low-density hot medium, which likely represents
the interior of a superbubble created by the co-existing OB association.
\end{enumerate}

We thank the referee Leisa Townsley for a thorough review of this work and
for useful comments that led to various improvements in the presentation
of the paper. We also thank Yang Su and Jiang-Tao Li for technical assistance.
Y.C.\ acknowledges support from NSFC grants 10073003 and 10221001
and grant NKBRSF-G19990754 of the China Ministry of Science and Technology,
while Q.D.W.\ acknowledges the support from NASA/CXC under the grant NAG5-3073A.
This publication makes use of data products from the Two Micron All Sky Survey,
which is a joint project of the University of Massachusetts and the Infrared
Processing and Analysis Center/California Institute of Technology,
funded by the National Aeronautics and Space Administration and the National
Science Foundation.
We also acknowledge the use of the SIMBAD and NED databases.


\begin{references} 
\reference{} Amato, E., Salvati, M., Bandiera, R., Pacini, F., 
 \& Woltjer, L.\ 2000, \aap, 359, 1107
\reference{} Bocchino, F.\ \& Bykov, A. M.\ 2001, \aap, 376, 248
\reference{} Borkowski, K.\ J., Lyerly, W.\ J., \& Reynolds, S.\ P.\
 2001, \apj, 548, 820
\reference{} Bucciantini, N.,  Amato, E., \& Del Zanna, L. 2005, A\&A, 434, 189
\reference{} Chevalier, R.\ A., 2000, \apjl, 539, 45
\reference{} Chu, Y.-H., Kennicutt, R.\ C., Jr., Schommer, R.\ A.,
 \& Laff, J.\ 1992, AJ, 103, 1545
\reference{} Crawford, F., McLaughlin, M., Johnston, S., Romani, R.,
 \& Sorrelgreen, E.\ 2005, AdSpR, 35, 1181
\reference{} Davis, J.\ E.\ 2001, \apj, 562, 575
\reference{} Dennerl, K., et al. 2001, \aap, 365, L202
\reference{} Fazio, G.~G., et al.\ 2004, \apjs, 154, 10
\reference{} Feigelson, E., et al. 2002, ApJ, 574, 258
\reference{} Frail, D. A., Giacani, E. B., Goss, W. M., \& Dubner, G.\
  1996, \apjl, 464, L165
\reference{} Gaensler, B.\ M., Chatterjee, S., Slane, P.\ O.,
 van der Swaluw, E., Camilo, F., Hughes, J.\ P.\ 2006, \apjl, in press,
 (astro-ph/0601304)
\reference{} Gotthelf, E. V., 2003, \apj, 591, 361
\reference{} Gotthelf, E. V. \& Wang, Q. D.  2000, \apjl, 532, 117
\reference{} Heger, A., Fryer, C.\ L., Woosley, S.\ E., Langer, N.,
 Hartmann, D.\ H.\ 2003, \apj, 591, 288
\reference{} Helfand, D. J., Gotthelf, E.V., \& Halpern, J.\ P.
  et al. 2001, \apj, ApJ, 556, 380
\reference{} Henize, K, G. 1956, \apjs, 2, 315
\reference{} Hester, J. J., Mori, K., Burrows, D., Gallagher, J. S.,
 Graham, J. R., Halverson, M., Kader, A., Michel, F. C., Scowen, P.\
 2002, \apjl, 577, L49
\reference{} Hwang, U., Holt, S. S., \& Petre, R.\ 2000, \apjl, 537, L119
\reference{} Jerius, D., Donnelly, R.\ H., Tibbetts, R.\ J., Edgar, R.\ J.,
 Gaetz, T.\ J., Schwartz, D.\ A., Speybroeck, L.\ P., \& Zhao, P.\
 2000, Proc.\ SPIE, 4012, 17
\reference{} Kaspi, V.\ M., Gotthelf, E.\ V., Gaensler, B.\ M.,
 \& Lyutikov, M.\ 2001, \apjl, 562, L163
\reference{} Komissarov, S.\ S.\ \& Lyubarsky, Y.\ E.\ 2003, \mnras, 344, L93
\reference{} Lazendic, J.\ S., Dickel, J.\ R., Haynes, R.\ F.,
 Jones, P.\ A., White, G.\ L.\ 2000, \apj, 540, 808
\reference{} Lu, F.\ J., Wang, Q.\ D., Aschenbach, B., Durouchoux, P.,
 \& Song, L.\ M.\ 2002, \apjl, 568, L49
\reference{} Lucke, P. B., \& Hodge, P. W. 1970, AJ, 75, 171
\reference{} Marshall, F.E., Gotthelf, E.V., Zhang, W., Middleditch, J.,
 \& Wang, Q. D. 1998, \apjl, 499, 179
\reference{} Marshall, F.E., Gotthelf, E.V., Middleditch, J., Wang, Q. D.,
 \& Zhang, W. 2004, \apj, 603, 682
\reference{} Matheson, H.\ \& Safi-Harb, S.\ 2005, AdSpR, 35, 1099.
\reference{} Mignani, R.\ P., Pulone, L., Iannicola, G., Pavlov, G.\ G.,
 Townsley, L., \& Kargaltsev, O.\ Y.\ 2005, \aap, 431, 659
\reference{} Mori, K; Burrows, D.\ N., Hester, J.\ J., Pavlov, G.\ G.,
 Shibata, S., Tsunemi, H.\ 2004, \apj, 609, 186
\reference{} Morrison, R., \& McCammon, D., 1983, \apj, 270, 119
\reference{} Ng, C.-Y.\ \& Romani, R.\ W., 2004, \apj, 601, 479
\reference{} Olbert, C.\ M., Clearfield, C.\ R., Williams, N.\ E.,
 Keohane, J.\ W., \& Frail, D.\ A., 2001, \apjl, 554, L205
\reference{} Oskinova, L. M.\ 2005, \mnras, 361, 679
\reference{} Park, S., Roming, P. W. A., Hughes, J. P., Slane, P. O.,
 Burrows, D. N., Garmire, G. P., \& Nousek, J. A., 2002, \apjl, 564, L39
\reference{} Petre, R., Kuntz, K.\ D., \& Shelton, R.\ L.\ 2002, \apj, 579, 404
\reference{} Reynolds, S.\ P.\ \& Chevalier, R.\ A.\ 1984, \apj, 278, 630
\reference{} Russell, S. C. \& Dopita, M. A.\ 1992, \apj, 384, 508
\reference{} Safi-Harb, S., Harrus, I. M., Petre, R., Pavlov, G. G.,
 Koptsevich, A. B., \& Sanwal, D.\ 2001, \apj, 561, 308
\reference{} Schild, H.\ \& Testor, G.\ 1992, \aaps, 92, 729
\reference{} Skrutskie, M.\ F., Cutri, R.\ M., Stiening, R.,
 Weinberg, M.\ D., Schneider, S., Carpenter, J.\ M., Beichman, C.,
 Capps, R., Chester, T., Elias, J., and 21 coauthors, 2006, \aj, 131, 1163
\reference{} Slane, P., Chen, Y, Schulz, N., Seward, F. D.,
 Hughes, J. P., \& Gaensler, B. M.\ 2000, \apjl, 533, L29
\reference{} Smith, D.\ A.\ \& Wang, Q.\ D.\ 2004, \apj, 611, 881
\reference{} Spitzer, L., Jr.\ 1982, \apj, 262, 315
\reference{} Sun, M., Wang, Z.-R., \&  Chen, Y.\ 1999, \apj, 511, 274
\reference{} Tang, S.,\ \& Wang, Q.\ D.\ 2005, \apj, 628, 205
\reference{} Thielemann, F.-K., Nomoto, K., \& Hashimoto, M.-A.\
1996, \apj, 460, 408
\reference{} Torii, K., Slane, P. O., Kinugasa, K., Hashimotodani, K.,
 \& Tsunemi, H.\  2000, \pasj, 52, 875
\reference{} Townsley, L.\ K., Broos, P.\ S., Feigelson, E.\ D.,
 Brandl, B.\ R., Chu, Y.-H., Garmire, G.\ P., Pavlov, G.\ G.\
 2006, \aj, 131, 2140
\reference{} van der Swaluw, E.\  2004, AdSpR, 33, 475
\reference{} van der Swaluw, E., Achterberg, A., Gallant, Y.\ A.,
 Downes, T.\ P., \& Keppens, R.\  2003, \aap, 397, 913
\reference{} van der Swaluw, E., Downes, T.\ P., \& Keegan, R.\ 2004,
 \aap, 420, 937
\reference{} Wang, Q. D. 2004, ApJ, 612, 159
\reference{} Wang, Q.\ D., \& Gotthelf, E. V. 1998, \apj, 494, 623 (WG98)
\reference{} Wang, Q.\ D., Gotthelf, E. V., Chu, Y.-H., \& Dickel, J. R.
 2001, \apj, 559, 275
\reference{} Wang, Q.\ D., Li, Z.-Y., \& Begelman, M. C. 1993, Nature, 364, 127
\reference{} Warren, J.\ S., Hughes, J.\ P., \& Slane, P.\ O.\ 2003, \apj,
 583, 260
\reference{} Warwick, R.\ S.\ et al.\ 2001, \aap, 365, L248
\reference{} Weisskopf, M.\ C., Hester, J. J., Tennant, A.\ F., Elsner, R.\ F.,
 Schulz, N.\ S., Marshall, H.\ L., Karovska, M., Nichols, J.\ S.,
 Swartz, D.\ A., Kolodziejczak, J.\ J., \& O'Dell, S.\ L., 2000, \apjl, 536, L81
\reference{} Woosley, S.\ E. \& Weaver, T.\ A.\ 1995, \apjs, 101, 181
\end{references}
\end{document}